\documentclass[]{aastex}

\usepackage{epsfig,amsfonts,amsmath,amssymb,graphicx,morefloats,appendix,tensor,gensymb}
\usepackage{color}
\usepackage{hyperref}
\newcommand\vsini{\ifmmode{v\sin{i_\star}}\else $v\sin{i_\star}$\fi}
\newcommand\sini{\ifmmode{\sin{i_\star}}\else $\sin{i_\star}$\fi}

\newcommand\mysim{\mathord{\sim}}

\newcommand{\rffigl}[1]{Figure~\ref{fig:#1}}

\newcommand{\rfsecl}[1]{\mbox{Section \ref{sec:#1}}}

\newcommand{\rftabl}[1]{Table~\ref{tab:#1}}

\newcommand{\fldg}{\hbox{\sc fld-g}}
\newcommand{\intg}{\hbox{\sc int-g}}
\newcommand{\vlg}{\hbox{\sc vl-g}}

\begin{document}

\title{Uniform Forward-Modeling Analysis of Ultracool Dwarfs.\\III. Late-M and L Dwarfs in Young Moving Groups, the Pleiades, and the Hyades}
\shorttitle{Uniform Forward-modeling Analysis of Ultracool Dwarfs. III.}

\author[0000-0002-6903-9080]{Spencer A. Hurt}
\affiliation{Department of Physics \& Astronomy, University of Wyoming, Laramie, WY 82071, USA}

\author[0000-0003-2232-7664]{Michael C. Liu}
\affiliation{Institute for Astronomy, University of Hawai‘i, 2680 Woodlawn Drive, Honolulu, HI 96822, USA}

\author[0000-0002-3726-4881]{Zhoujian Zhang}\thanks{NASA Sagan Fellow} 
\affiliation{Department of Astronomy \& Astrophysics, University of California, Santa Cruz, 1156 High St, Santa Cruz, CA 95064, USA}
\affiliation{Department of Astronomy, The University of Texas at Austin, 2515 Speedway, Stop C1400, Austin, TX 78712, USA}

\author[0000-0001-6041-7092]{Mark Phillips}
\affiliation{Institute for Astronomy, University of Hawai‘i, 2680 Woodlawn Drive, Honolulu, HI 96822, USA}

\author[0000-0003-0580-7244]{Katelyn N. Allers}
\affiliation{Department of Physics \& Astronomy, Bucknell University, Lewisburg, PA 17837, USA}

\author[0000-0003-2440-7350]{Niall R. Deacon} 
\affiliation{Max Planck Institute for Astronomy, Königstuhl 17, 69117 Heidelberg, Germany}

\author{Kimberly M. Aller}
\affiliation{Institute for Astronomy, University of Hawaii at Manoa, Honolulu, HI 96822, USA}

\author[0000-0003-0562-1511]{William M. J. Best} 
\affiliation{University of Texas at Austin, Department of Astronomy, 2515 Speedway C1400, Austin, TX 78712, USA}

\shortauthors{Hurt et al.}

\begin{abstract}
We present a uniform forward-modeling analysis of 90 late-M and L dwarfs in nearby young ($\mysim10-200$ Myr) moving groups, the Pleiades, and the Hyades using low-resolution ($R\approx150$) near-infrared ($0.9-2.4$ $\mathrm{\mu m}$) spectra and the BT-Settl model atmospheres. We derive the objects' effective temperatures, surface gravities, radii, and masses by comparing our spectra to the models using a Bayesian framework with nested sampling and calculate the same parameters using evolutionary models. Assuming the evolutionary-based parameters are more robust, our spectroscopically inferred parameters from BT-Settl exhibit two types of systematic behavior for objects near the M-L spectral type boundary. Several objects are clustered around $T_\mathrm{eff} \approx 1800$ K and $\log g\approx5.5$ dex, implying impossibly large masses ($150-1400$ $M_\mathrm{Jup}$), while others are clustered around $T_\mathrm{eff}\gtrsim3000$ K and $\log g\lesssim3.0$ dex, implying non-physical low masses and unreasonably young ages. We find the fitted BT-Settl model spectra tend to overpredict the peak $J$ and $H$-band flux for objects located near the M-L boundary, suggesting the dust content included in the model atmospheres is insufficient to match the observations. By adding an interstellar medium-like reddening law to the BT-Settl model spectra, we find the fits between models and observed spectra are greatly improved, with the largest reddening coefficients occurring at the M-L transition. This work delivers a systematic examination of the BT-Settl model atmospheres and constitutes the largest spectral analysis of benchmark late-M and L-type brown dwarfs to date. 
\end{abstract}

\section{Introduction}
\label{sec:introduction}

Brown dwarfs ($\approx$13 to $\approx$70 $M_\mathrm{Jup}$; \citealt{2011ApJ...727...57S, 2017ApJS..231...15D}) have many properties that overlap with those of directly-imaged planets, including effective temperature and surface gravity. Atmospheric characterization of imaged planets through emission spectroscopy (e.g., \citealt{2016AJ....152..203L, 2017A&A...603A..57S}) is often challenging, given that most are closely separated from their much brighter host stars. In contrast, non-irradiated brown dwarfs are readily observed via high-quality emission spectroscopy, offering the opportunity to study physical and chemical processes similar to those at play in the atmospheres of gas-giant planets. Observations of brown dwarfs can therefore provide valuable insight into the atmospheric properties and formation pathway of planetary-mass objects \citep[e.g.,][]{2001RvMP...73..719B, 2008ApJ...683.1104F, 2015ARA&A..53..279M}. 

Atmospheric studies of brown dwarfs have traditionally relied on forward modeling (e.g., \citealt{2000ApJ...541..374S, 2001ApJ...556..373G, 2006ApJ...639.1095B, 2008ApJ...678.1372C, 2009ApJ...702..154S, 2010ApJS..186...63R, 2011ApJ...743...50C, 2011ApJ...740..108L, 2016MNRAS.455.1341M, 2014A&A...562A.127B, 2018A&A...618A..63B, 2020ApJ...891..171Z}), where the observed spectra are compared to a grid of pre-computed theoretical spectra based on several input physical parameters (typically including effective temperature, surface gravity, and metallicity) and self-consistent assumptions (such as radiative-convective equilibrium, equilibrium and disequilibrium chemistry, and cloud properties). \cite{2021ApJ...916...53Z} forward-modeled 3 benchmark late-T dwarfs, evaluating the performance of the new cloud-free Sonora Bobcat model atmospheres \citep{2021ApJ...920...85M}. \cite{2021ApJ...921...95Z} extended this analysis to derive the atmospheric properties of 55 field late-T dwarfs and conducted a systematic examination of Sonora Bobcat models. Together, they found that the spectroscopically inferred metallicities from the Sonora Bobcat models are likely underestimated by 0.3-0.4 dex, leading to implausibly young ages. The inferred model-based effective temperatures of objects with spectral types T8 or later are likely overestimated by 50 to 200 K, leading to unphysically small masses. Additionally, they identified systematic differences in the observed and the fitted model spectra, which tend to overpredict the fluxes in $J$-band ($1.18-1.35$ $\mathrm{\mu m}$). Such analyses are crucial towards interpreting the parameters returned by forward models and can help improve future model atmospheres.  

In this paper, we extend the work done by \cite{2021ApJ...916...53Z} and \cite{2021ApJ...921...95Z} to late-M and L dwarfs by conducting a uniform forward-modeling analysis to 90 M5-L9 young ultracool dwarfs using the cloudy BT-Settl atmospheric models \citep{2012RSPTA.370.2765A}. While previous studies have used limited samples to study the performance of BT-Settl for M dwarfs \citep[e.g.,][]{2018A&A...610A..19R} and L dwarfs \citep[e.g.,][]{2014MNRAS.439..372M, 2016ApJ...830...96H}, none have evaluated these models across the full M-L transition boundary. With the largest spectral sample of brown dwarfs with known ages to date, this work aims to provide context for the predictive power of other cloudy models developed for planetary-mass objects and brown dwarfs as well as assess the reliability of these widely used BT-Settl models. In \rfsecl{targets}, we summarize the targets in our sample, the corresponding observations, and our spectral type and gravity classification analysis. We then review the BT-Settl atmospheric models in \rfsecl{bt-settl} and construct our forward modeling framework in \rfsecl{atmosphericanalysis}. In addition to fitting the spectra using atmospheric models, we also derive the physical properties of benchmark objects in our sample using their ages and bolometric luminosities with evolutionary models in \rfsecl{evolutionaryanalysis}. Using these results, we examine the performance of the BT-Settl models in \rfsecl{benchmarking}. In \rfsecl{extinction}, we demonstrate that extinction laws for interstellar dust improve fits between observed and model spectra. We then conclude our results in \rfsecl{conclusions}.

\begin{figure}
    \centering
    \includegraphics[width=\linewidth]{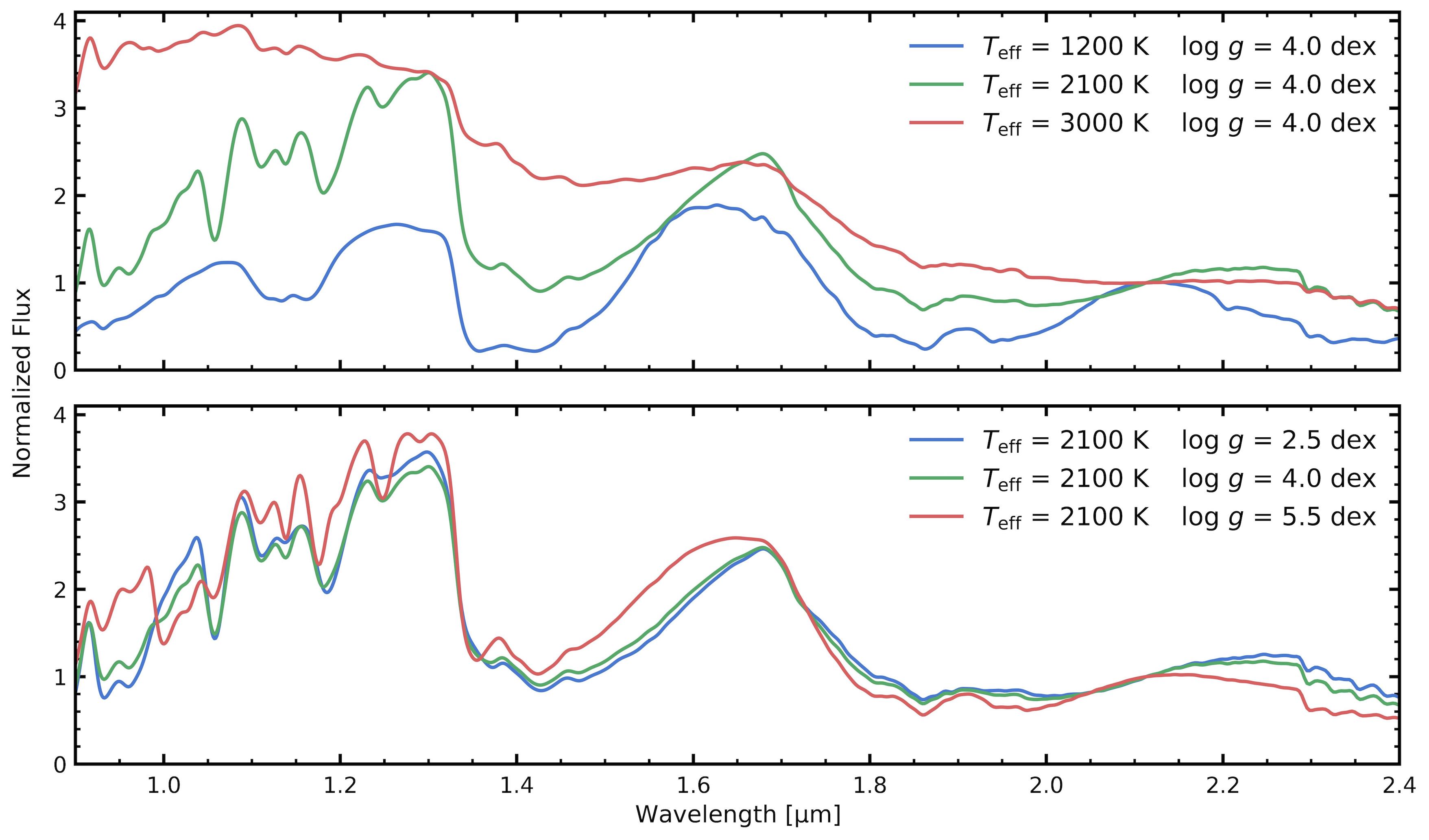}
    \caption{Comparisons between cloudy BT-Settl model spectra with varying $T_\mathrm{eff}$ (top panel) and $\log g$ (bottom panel). All spectra have been downgraded to the wavelength-dependent resolution of the 0.5" slit of the SpeX prism mode ($R\approx80-250$) and are normalized using their fluxes at 2.15 $\mathrm{\mu m}$.}
    \label{fig:examples}
\end{figure}

\section{Sample and Observations}
\label{sec:targets}

\subsection{Pleiades Objects}
\label{sec:sample-pleiades}
The Pleiades provide a rich locale for studies of substellar evolution, given the many photometric, spectroscopic, and astrometric searches for low-mass members over the decades (e.g., \citealt{2015A&A...577A.148B} and references therein). We selected from the literature a sample of 26 late-M and L~dwarf Pleiades members for near-IR spectroscopic characterization (\rftabl{sample}). Most of the objects have already been spectroscopically confirmed at optical and/or near-IR wavelengths \citep[e.g.,][]{1998AA...336..490B,2000ApJ...543..299M,2010A&A...519A..93B}, while a few had only been identified as photometric candidates (e.g., \citealt{2012MNRAS.422.1495L}). Most of these objects have parallactic distances from {\em Gaia} EDR3, and 9 of them have moving-cluster distances from \citet{2017A&A...598A..48G}. One object (Roque~25) does not have a distance estimate in the literature, so we adopt the cluster distance of 134$\pm$9~pc from \cite{2018ApJ...856...23G}.

After we obtained spectra of this sample, 3 of the objects (SI2M-28, CFHT-PL-6, CFHT-PL-15) turned out to be non-members as shown both by the analysis of \citet{2015A&A...577A.148B} based on a probabilistic model and our membership analysis using the BANYAN~$\Sigma$ algorithm of \citet{2018ApJ...856...23G} with the latest astrometry in the {\tt UltracoolSheet} described below. In brief, the EDR3 parallaxes for SI2M-28 and CFHT-PL-15 place them behind the cluster and the parallax for CFHT-PL-6 places it in the foreground. For the remaining 23 objects, 22 of them have very high membership probabilities from the \citet{2015A&A...577A.148B} analysis, with 21 of them having $>$99.6\% probability and the remaining one having 95.3\% (for HHJ~3). We also ran the BANYAN~$\Sigma$ algorithm on these objects. Most have high ($\gtrsim$80\%) membership probabilities. For the two objects that do not (Roque~7 and UGCS~J034815.64+255008.9), \citet{2015A&A...577A.148B} report $>$99.9\% membership probability, and these objects' near-IR spectra indicate low-gravity, so we consider these to be Pleiades members.

\subsection{Young Moving Group and Hyades Members}
\label{sec:sample-ymg}
We identified a sample of young moving group (YMG) and Hyades candidates using the {\tt UltracoolSheet} \citep{best_william_m_j_2020_4169085}, a catalog of 3000+ ultracool dwarfs developed from compilations by \cite{2012ApJS..201...19D}, \cite{2013Sci...341.1492D}, \cite{2016ApJ...833...96L}, \cite{2018ApJS..234....1B}, and \cite{2021AJ....161...42B}. The {\tt UltracoolSheet} includes an analysis of moving group membership for all objects using the BANYAN~$\Sigma$ algorithm \citep{2018ApJ...856...23G} and the best available astrometry. We selected objects with infrared spectral types L9 or earlier and membership probabilities greater than 90\%. We then identified those objects with existing low-resolution ($R\approx50-250$) near-IR spectra taken with the SpeX spectrograph \citep{2003PASP..115..362R} located at the NASA Infrared Telescope Facility (IRTF). Most of the spectra come from the SpeX Prism Library \citep{2014ASInC..11....7B, 2017ASInC..14....7B}, with 11 objects having spectra from our own observations (described below). This sample comprises 59 YMG and 5 Hyades objects, all of which can be found in \rftabl{sample}.

One object in our sample, PSO J243.9421+67.2075, was previously unpublished and was discovered by our group through a search for very red L~dwarfs based on Pan-STARRS~1 photometry, similar to the process described in \cite{2013ApJ...777L..20L}.

At the start of this work, three objects in our YMG sample, 2MASS J03264225-2102057, WISEA J235422.31-081129.7, and PSO J243.9421+67.2075, were thought to be members of either the AB Doradus (ABDMG) and Beta Pictoris (BPMG) moving groups, but updated astrometry has revealed that all three likely field objects. Our final sample includes 56 YMG objects (belonging to 7 total moving groups), 5 Hyades objects, 23 Pleiades objects, and 6 field objects. Details about the moving groups are provided in \rftabl{ymg}.

\begin{figure}
    \centering
    \includegraphics[width=\linewidth]{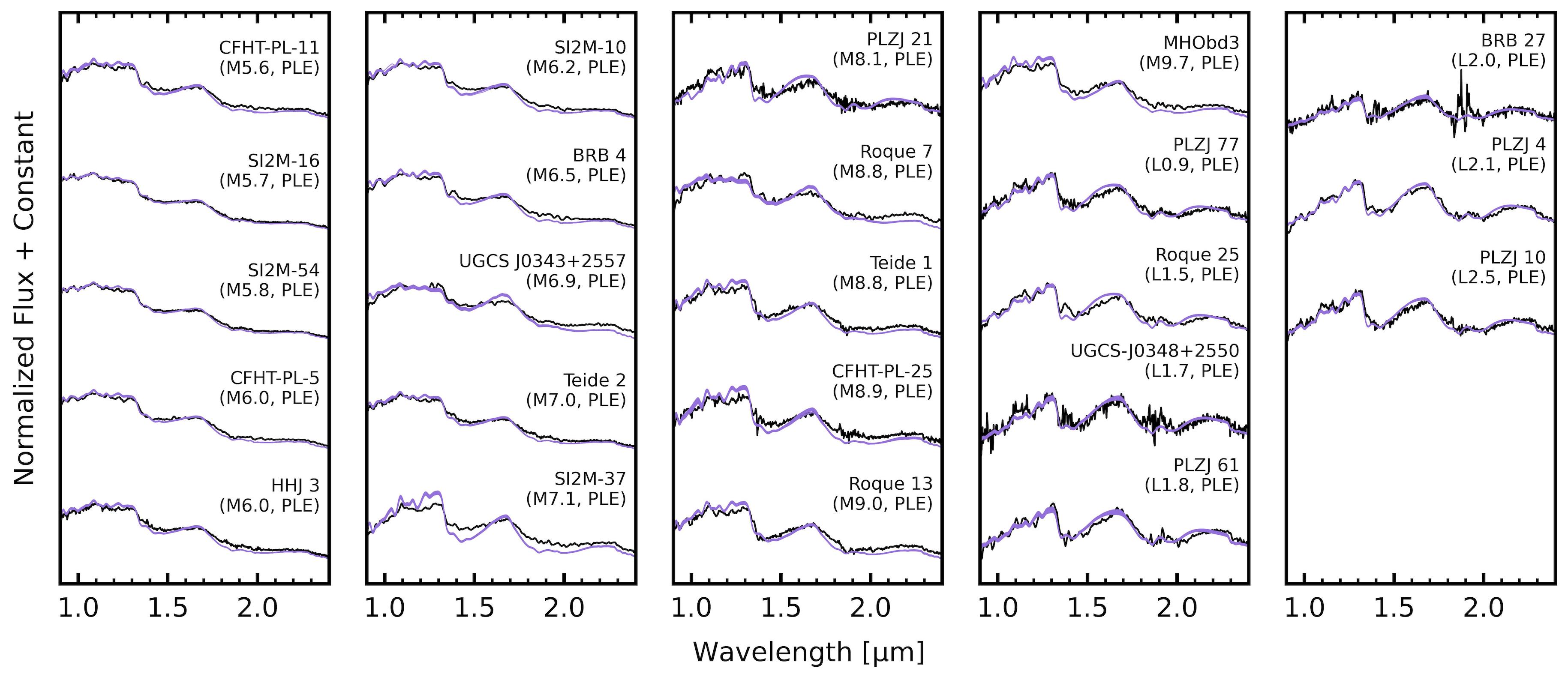}
    \includegraphics[width=\linewidth]{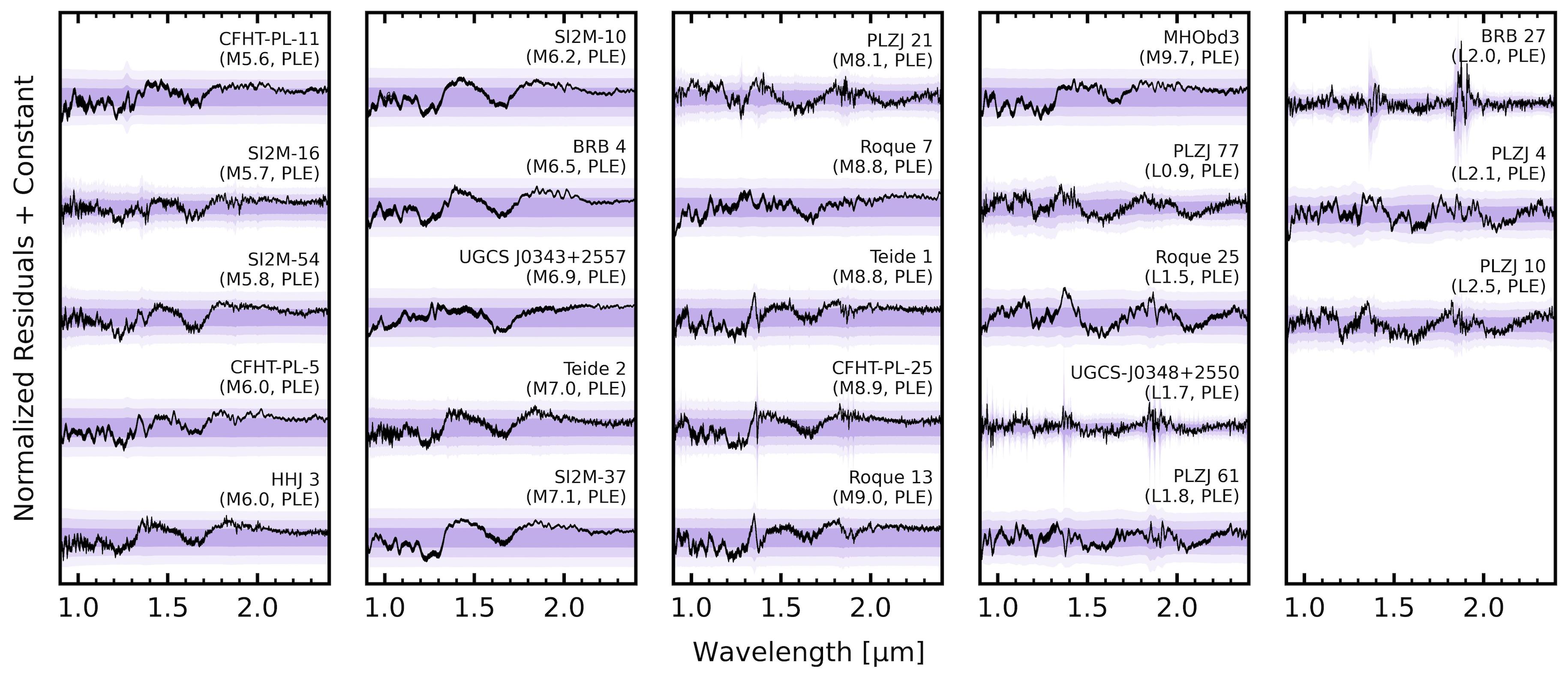}
    \caption{The top panel shows the observed spectra as black lines plotted against 250 BT-Settl models (purple lines) drawn from the posteriors for each object located in the Pleiades. These posteriors come directly from the formal atmospheric model fitting and do not include systematic uncertainties. The bottom figure shows the corresponding residuals as black lines along with the 1$\sigma$, 2$\sigma$, and 3$\sigma$ dispersions (the measurement uncertainties and $\epsilon_\mathrm{med}$ added in quadrature) as purple shadows.}
    \label{fig:allspectra_pleiades}
\end{figure}

\begin{figure}
    \centering
    \includegraphics[width=\linewidth]{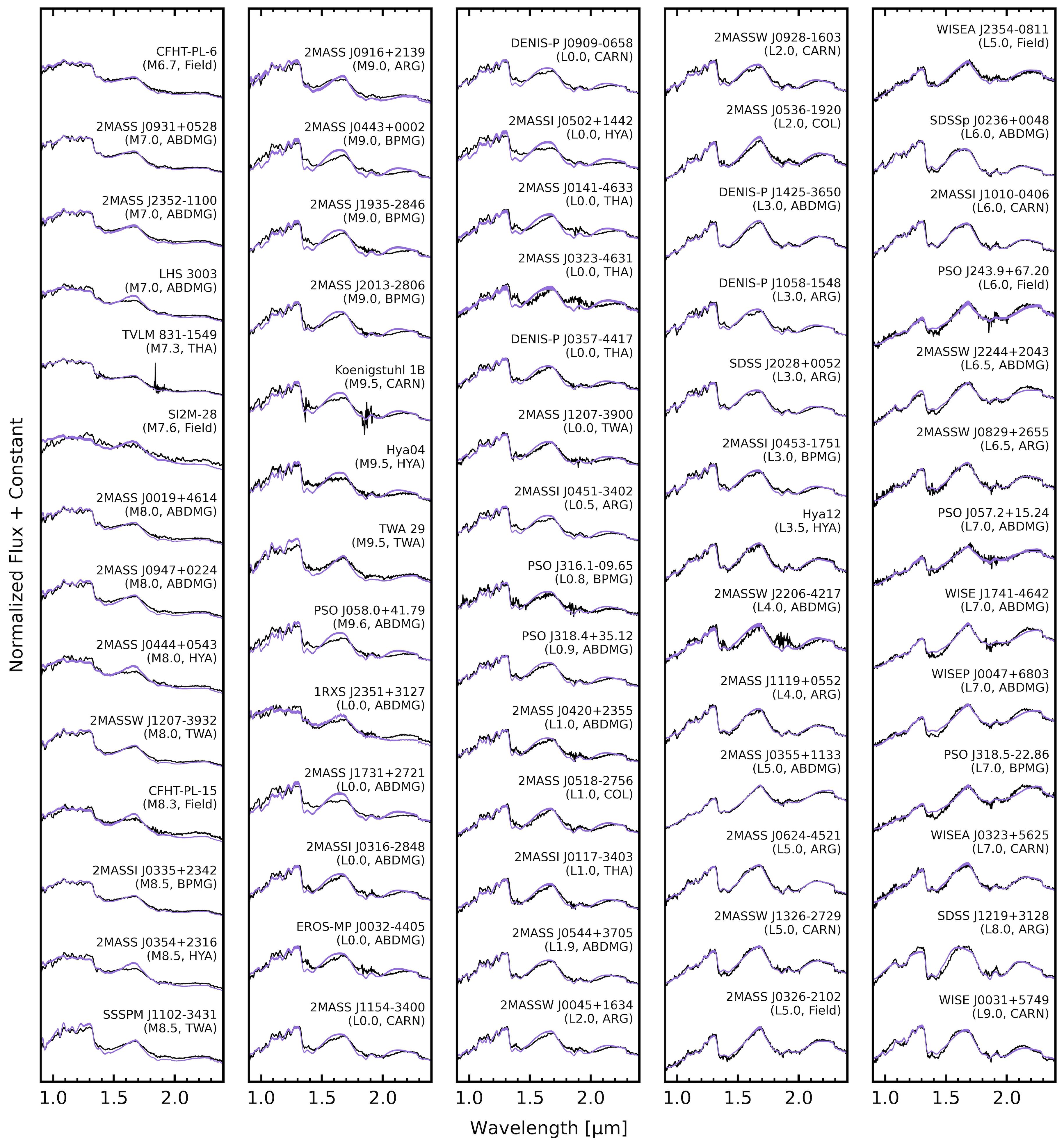}
    \caption{Observed spectra are shown as black lines plotted against 250 BT-Settl models (purple lines) drawn from the posteriors for each YMG and Hyades object in our sample. These posteriors come directly from the formal atmospheric model fitting and do not include systematic uncertainties.}
    \label{fig:allspectra_nopleiades}
\end{figure}

\begin{figure}
    \centering
    \includegraphics[width=\linewidth]{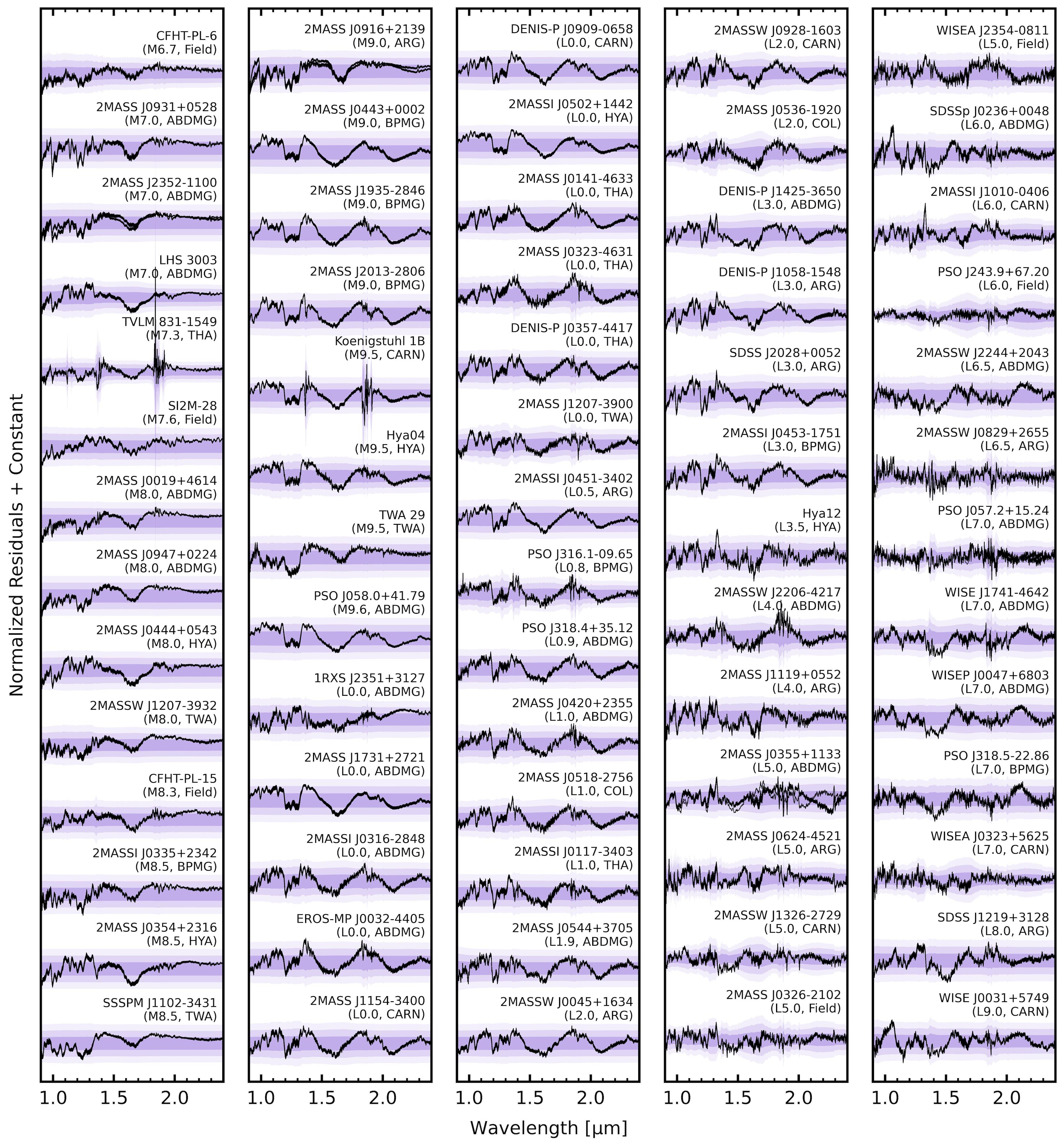}
    \caption{The residuals (data$-$model) between the observed spectra and 250 BT-Settl models drawn from the posteriors for each YMG and Hyades object in our sample are shown in black. These posteriors come directly from the formal spectral fitting and do not include systematic uncertainties. The 1$\sigma$, 2$\sigma$, and  3$\sigma$ dispersions (corresponding to the measurement uncertainties and $\epsilon_\mathrm{med}$ added in quadrature) are shown as purple shadows.}
    \label{fig:allresiduals_nopleiades}
\end{figure}

\subsection{IRTF/SpeX Observations}

We obtained low-resolution near-IR (0.8--2.5~\micron) spectra for 8 YMG objects, 1 Hyades object, 23 Pleiades objects, and 5 field objects from the NASA Infrared Telescope Facility (IRTF) on the summit of Maunakea, Hawaii. We used the SpeX spectrograph \citep{2003PASP..115..362R} in prism mode with either the 0.5\arcsec\ or 0.8\arcsec\ slit, which provided an average spectral resolution ($R\equiv\Delta\lambda/\lambda$) of $\approx$150 or $\approx$100, respectively. We dithered in an ABBA pattern on the science target to enable sky subtraction during the reduction process. For each epoch, we obtained calibration frames (arcs and flats) and observed an A0~V star contemporaneously for telluric calibration. All spectra were reduced using the Spextool software package \citep{2003PASP..115..389V,2004PASP..116..362C}. \rftabl{spex_observing} summarizes our observations. 

We determined near-IR spectral types and gravity classifications using the widely-used \cite{2013ApJ...772...79A} system, developed precisely for low ($R\approx150$) and medium ($R\approx1000$) resolution spectra obtained by SpeX. Spectral types are established using both visual classification (i.e., the traditional comparison to predefined standard objects) and quantitative classification (i.e., using measurement of spectral indices). The near-IR spectral standards are drawn from \cite{2010ApJS..190..100K}. The spectral indices for low-resolution data are chosen to measure the depth of the near-IR water  bands, which can be well-measured even at low resolution. Polynomial relations between the near-IR index values and spectral types are established using optical spectral types. Similarly, gravity classifications (\vlg, \intg, and \fldg) are established using gravity-sensitive indices, which measure the strength of the FeH, VO and \ion{K}{1} lines as well as the shape of the $H$-band continuum. A few objects have \fldg\ classifications residing in ABDMG, CARN, and ARG. For the first two groups, such classifications are not surprising, given previous evidence that objects of $\approx$150~Myr may show such spectral signatures (e.g., \citealt{2016PhDT.......189A}). ARG is rather younger, though there are instances of a candidate members also having such a gravity class (\citealt{2015ApJS..219...33G}, \citealt{2016ApJ...833...96L}), which might be due to interlopers in the group membership or reflecting the imprecision of such classification at low spectral resolution (e.g., see Figure 21 of \citealt{2023arXiv230903082S}). The spectral types and gravity classes derived from our SpeX spectra are given in \rftabl{sample}. For the Pleiades objects, \rftabl{spex_observing} summarizes previous literature measurements of their optical and near-IR spectral types, which agree well with our determinations given that differences of $\pm$1 subclass are common for late-M and L dwarfs.

For both our spectra and those from the SpeX Prism Library, the spectrum for each object was flux-calibrated using the corresponding $H_\mathrm{MKO}$ magnitude and the measurement uncertainties were inflated accordingly. The photometry and information about the spectra for our sample are given in \rftabl{photometry}.

\section{BT-Settl Atmospheric Models}
\label{sec:bt-settl}
Throughout this work, we use the CIFIST 2011-2015 BT-Settl model atmosphere\footnote{The model grid can be downloaded at \url{http://svo2.cab.inta-csic.es/theory/newov2/index.php}.} \citep{2012RSPTA.370.2765A, 2015A&A...577A..42B}, which is distinguished from other publicly available models by its treatment of clouds and dust in the atmospheres of cool objects. BT-Settl synthetic spectra are computed as a function of effective temperature ($T_\mathrm{eff}$) and surface gravity ($\log g$) while assuming solar metallicity ($Z=0$ dex, as defined by \citealt{2011SoPh..268..255C}). First, the abundance and size distribution of dust particles throughout the atmospheres were computed using the timescales of condensation, coalescence, mixing, and gravitational settling for 55 grain types (details of the solids are given in \citealt{2018A&A...620A.180R}). The wavelength-dependent opacities were then calculated line-by-line and fed to the PHOENIX code \citep{2001ApJ...556..357A}, which handled the radiative transfer assuming hydrostatic equilibrium and accounted for convection using mixing-length theory. Additionally, the BT-Settl model spectra account for non-equilibrium chemistry between molecules including CO, CH$_4$, CO$_2$, N$_2$, and NH$_3$. Compared to other versions of the BT-Settl models, the CIFIST 2011-2015 grid benefited from updated molecular line lists and utilized convective mixing lengths calibrated using multi-dimensional radiative hydrodynamics, as described by \cite{2015A&A...577A..42B}.

Most other 1D model atmospheres that account for dust include free parameters corresponding to the spatial distribution or mass density of clouds. For example, \cite{2008ApJ...689.1327S} use the sedimentation frequency ($f_\mathrm{sed}$), developed first by \cite{2001ApJ...556..872A}, where larger values correspond to rapid particle growth; in these instances, condensates more rapidly fall out of the atmosphere, resulting in optically thinner clouds. The petitCODE models \citep{2015ApJ...813...47M, 2017A&A...603A..57S} use a similar cloud sedimentation parameter. And while the Exo-REM models \citep{2015A&A...582A..83B,2017ApJ...850..150B} utilize the same microphysics as BT-Settl, they include a cloud fraction parameter ($f_\mathrm{cloud}$) to represent a mix of clear and cloudy regions in the atmosphere. Besides varying treatments of clouds, these models consider different lists of chemical species and cover different ranges of $T_\mathrm{eff}$ and $\log g$. BT-Settl notably spans a wide range of temperatures that includes both M and L dwarfs.

We use the full BT-Settl CIFIST 2011-2015 models with 446 grid points spanning $T_\mathrm{eff}$ of $1200-7000$~K (in intervals of either 50~K or 100~K) and $\log g$ of $2.5-5.5$~dex (in intervals of 0.5 dex). \rffigl{examples} presents several examples of BT-Settl spectra downgraded to the resolution of the SpeX prism 0.5" slit and illustrates how spectral morphology changes as a function of $T_\mathrm{eff}$ or $\log g$.

\begin{figure}
    \centering
    \includegraphics[width=\linewidth]{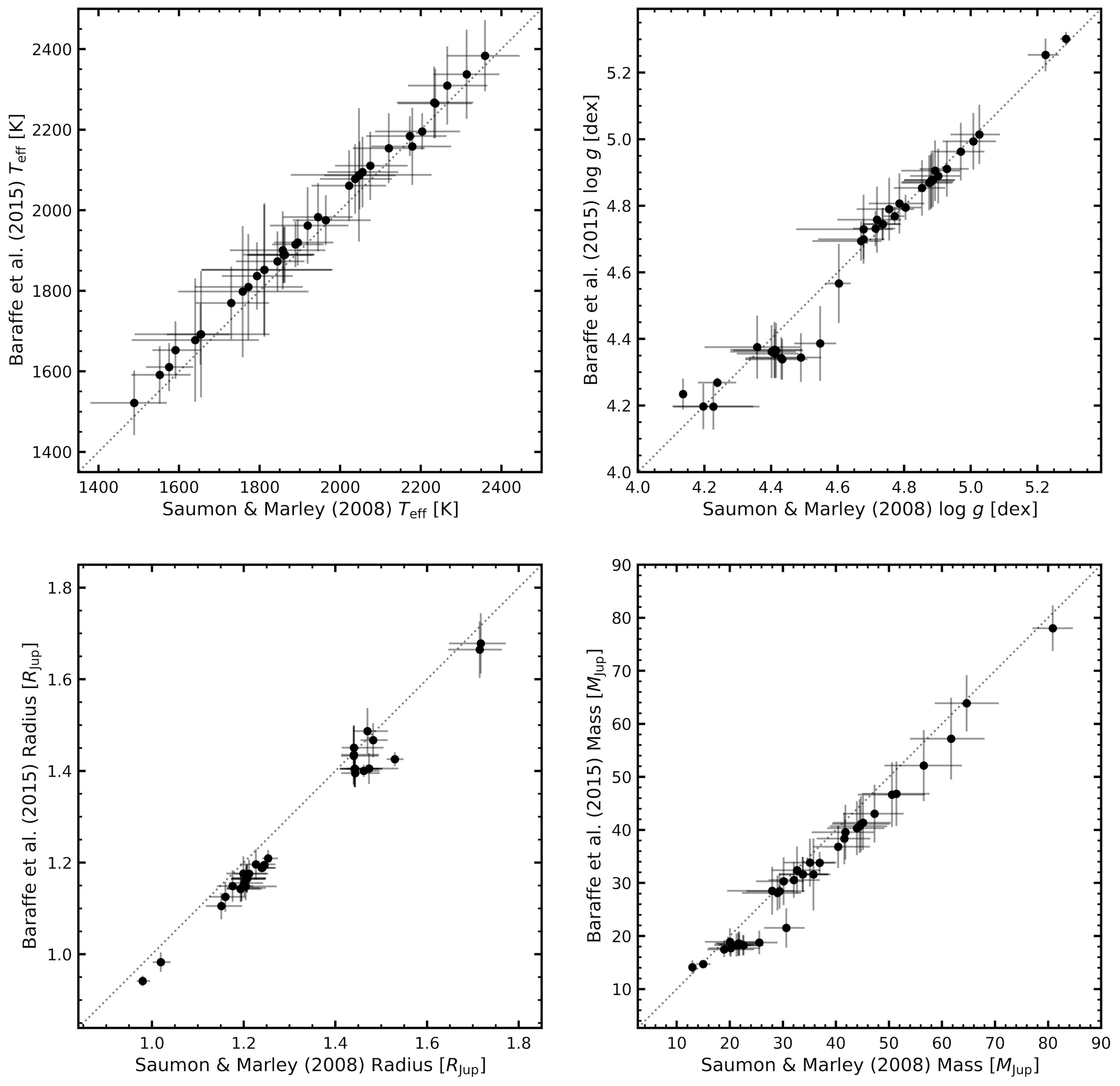}
    \caption{Comparison of physical parameters derived using the cloudy evolutionary models from \cite{2008ApJ...689.1327S} and \cite{2015A&A...577A..42B}. The top left panel compares the derived $T_\mathrm{eff}$, the top right compares $\log g$, the bottom left $R$, and the bottom right $M$. While the two models are nearly always consistent with each other, we see that the \cite{2015A&A...577A..42B} model returns median $T_\mathrm{eff}$ typically higher ($\sim$25 K) and median $R$ and $M$ typically lower ($\sim$0.04 $R_\mathrm{Jup}$ and $\sim$2 $M_\mathrm{Jup}$, respectively) than those returned by the \cite{2008ApJ...689.1327S} hybrid model.}
    \label{fig:evovevo}
\end{figure}

\section{Atmospheric Model Analysis}
\label{sec:atmosphericanalysis}

Forward-modeling analyses typically compare an observed spectrum with the synthetic spectra from a set of model atmospheres and derive the best-fit parameters using least-squares minimization. Because model grids tend to be coarsely spaced, linear interpolation between synthetic spectra is frequently used to estimate parameters between grid points. However, linear interpolation can lead to underestimated error bars on the derived physical parameters and may bias posteriors towards the original model grid points \citep[e.g.,][]{2014ApJ...794..125C, 2015ApJ...812..128C, 2020ApJ...891..171Z, 2021ApJ...921...95Z}. To mitigate these concerns, \cite{2021ApJ...916...53Z} and \cite{2021ApJ...921...95Z} used the Bayesian framework described by \cite{2015ApJ...812..128C}, which was implemented through the Python package {\tt Starfish} \citep{ian_czekala_2018_2221006}. Rather than linearly interpolate the model grid, {\tt Starfish} employs a spectral emulator that combines principle component analysis (PCA) with Gaussian processes (GPs) to calculate a probability distribution of interpolated model spectra for a given set of parameters, allowing interpolation uncertainties to be propagated into the inferred posteriors. Additionally, {\tt Starfish} includes a covariance matrix with off-diagonal components to account for correlated residuals among adjacent wavelength pixels associated with instrumental properties and modeling systematics.

Although the framework used by \cite{2021ApJ...916...53Z} and \cite{2021ApJ...921...95Z} can return more realistic posteriors for model atmospheres such as Sonora Bobcat, we find that our trained spectral emulator is unable to reconstruct the BT-Settl synthetic spectra across the entire grid. This is because the cloudy BT-Settl model atmospheres do not behave monotonically with effective temperature or surface gravity for low $T_\mathrm{eff}$ ($\leq1800$~K); consequently, the PCA coefficients do not vary smoothly with $T_\mathrm{eff}$ or $\log g$ and cannot be well-described by a GP. Therefore, our analysis does not utilize the {\tt Starfish} framework and instead relies on linear interpolation between grid points. To avoid under-reporting the error bars on our derived parameters, we follow previous works that adopt fractions of the model grid spacing as the final parameter uncertainties (e.g., \citealt{2007ApJ...667..537L, 2008ApJ...678.1372C, 2009ApJ...702..154S, 2021ApJ...921...95Z}).

\begin{figure}
    \centering
    \includegraphics[width=\linewidth]{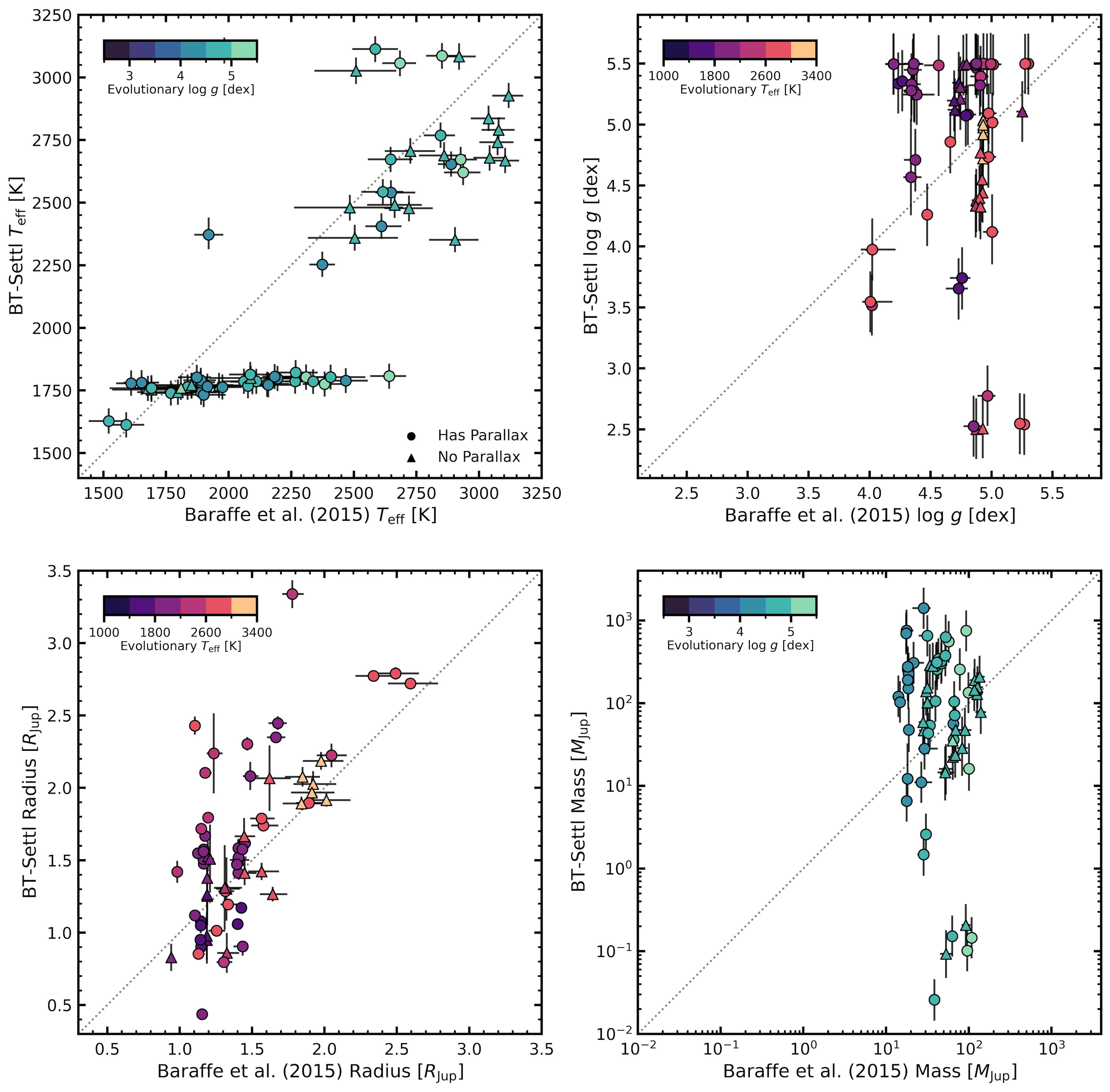}
    \caption{Physical parameters derived using BT-Settl model atmospheres versus those derived using the \cite{2015A&A...577A..42B} evolutionary models. The top left panel compares $T_\mathrm{eff}$, the top right compares $\log g$, the bottom left $R$, and the bottom right $M$. The points in the top left and bottom right panels are colored according to the surface gravity determined using the evolutionary model while the points in the top right and bottom left are colored according to effective temperature determined using the evolutionary model. Circular data points indicate an object that has a parallax measurement while triangular data points indicate otherwise.}
    \label{fig:btsettl-vs-baraffe}
\end{figure}

\begin{figure}
    \centering
    \includegraphics[width=\linewidth]{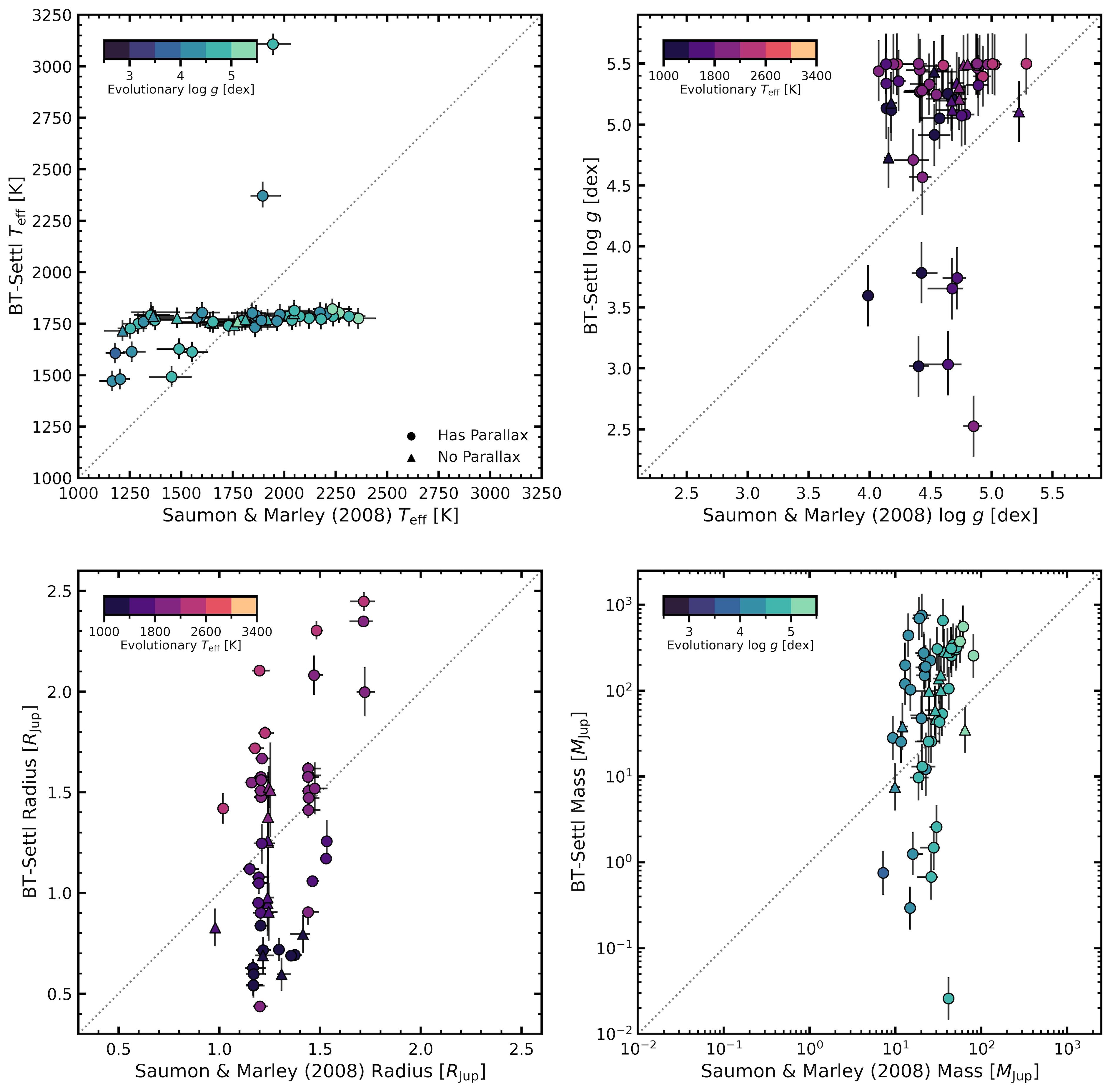}
    \caption{Physical parameters derived using BT-Settl model atmospheres versus those derived using the \cite{2008ApJ...689.1327S} evolutionary models. The top left panel compares $T_\mathrm{eff}$, the top right compares $\log g$, the bottom left $R$, and the bottom right $M$. The points in the top left and bottom right panels are colored according to the surface gravity determined using the evolutionary model while the points in the top right and bottom left are colored according to effective temperature determined using the evolutionary model. Circular data points indicate an object that has a parallax measurement while triangular data points indicate otherwise.}
    \label{fig:btsettl-vs-sm08}
\end{figure}

We begin our analysis by downgrading the resolution of each synthetic spectrum in the BT-Settl grid using Gaussian profiles corresponding to the 0.3", 0.5", or 0.8" slit SpeX prism observing modes, returning three model grids matching the three different resolutions of SpeX prism spectra. This convolution process accounts for the wavelength-dependent resolution of the prism mode \citep{2003PASP..115..362R}. We then create models for each slit size that take $\log T_\mathrm{eff}$ and $\log g$ as inputs and linearly interpolate over all 446 degraded BT-Settl spectra to return predicted fluxes at a given set of wavelengths. Our models also adopt $\log\Omega$ as a parameter and rescale the interpolated spectra by a factor of $\Omega$. Note that this scale factor is equivalent to the solid angle occupied by an object ($\Omega=R^2/d^2$, where $R$ is the object radius and $d$ is the distance to the object from an observer). Given the low spectral resolution of our data, rotational broadening or Doppler shifting can be excluded without biasing the inferred atmospheric properties. The objects in our sample are located at close enough distances (7-145~pc) that the effects of interstellar extinction are also negligible.

When fitting these models to our observed spectra, we include a jitter term ($\xi$) that is added in quadrature to the spectral flux uncertainties to account for the instrumental noise and systematic differences between data and model predictions. The log-likelihood function for our models is given by
\begin{equation}
    ts\mathcal{L} = -\frac{1}{2}\sum_\lambda\frac{\left(F_\lambda\left(\log T_\mathrm{eff}, \log g, \log\Omega\right) - f_\lambda\right)^2}{\sigma_\lambda^2 + \xi^2} - \ln\sqrt{2\pi\left(\sigma_\lambda^2 + \xi^2\right)},
\end{equation}
where $F_\lambda$ is the model flux, $f_\lambda$ is the observed flux, and $\sigma_\lambda$ is the observed flux uncertainty at a wavelength of $\lambda$. The second term in our log-likelihood function acts as a penalty term for large $\xi$, ensuring that the jitter term does not become unphysically large, accounting for all variability in our data (e.g., \citealt{2018PASP..130d4504F}). Uninformative uniform priors are additionally placed on $\log T_\mathrm{eff}$ and $\log g$ to ensure that they do not fall beyond the values spanned by the BT-Settl grid. 

While Markov chain Monte Carlo (MCMC) processes like those implemented in {\tt emcee} \citep{2013PASP..125..306F} or {\tt zeus} \citep{karamanis2020ensemble, karamanis2021zeus} are often used for forward-modeling, we find that the likelihood-surfaces for the BT-Settl fits are often multi-modal and too complex to be well-explored by these algorithms. Further, they tend to be computationally intensive and slow to converge. Therefore, we use the nested sampling algorithm implemented in {\tt UltraNest} \citep{2016S&C....26..383B, 2019PASP..131j8005B, 2021JOSS....6.3001B} to explore the posteriors of each object in our sample. The fitted spectra and residuals for the Pleiades members are shown in \rffigl{allspectra_pleiades} while the fitted spectra and residuals for the remaining objects can be found in \rffigl{allspectra_nopleiades} and \rffigl{allresiduals_nopleiades}, respectively.

To account for the challenges encountered with linear interpolation discussed at the beginning of this section, we inflate the derived $T_\mathrm{eff}$ and $\log g$ posteriors for each object with random Gaussian noise centered at 0 and standard deviations equal to half the spacing between the BT-Settl grid points. Using the known distance to each object, we derive the radius posterior from the $\log\Omega$ posterior for each object. We then combine the surface gravity and radius posteriors to derive masses ($M$) for our sample. We additionally define the parameter $\epsilon_\mathrm{med}$ by dividing each $\xi$ posterior by the median flux of the observed spectrum, providing a reference point for how large the jitter is and how well the model fits the data (lower values indicate a better fit). The final posteriors for $T_\mathrm{eff}$, $\log g$, $R$, $M$, and $\epsilon_\mathrm{med}$ are listed in \rftabl{derived_btsettl}.

\begin{figure}[!t]
    \centering
    \includegraphics[width=\linewidth]{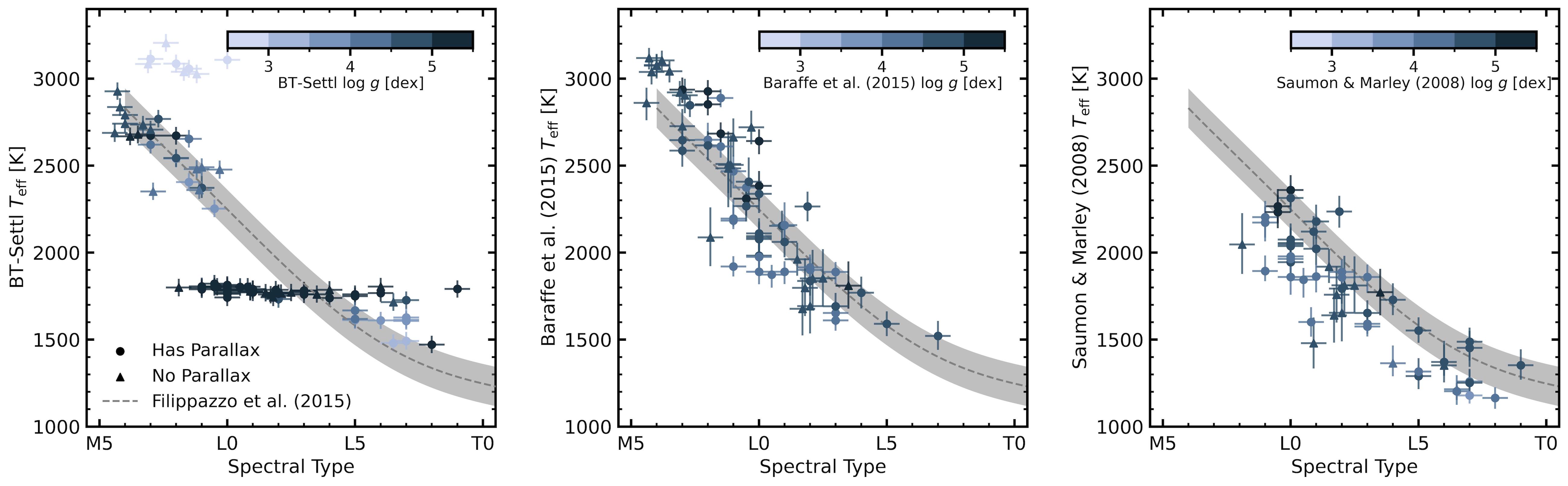}
    \caption{Derived effective temperature versus object spectral type using $T_\mathrm{eff}$ derived from BT-Settl (left panel), the \cite{2015A&A...577A..42B} evolutionary models (middle panel), and the \cite{2008ApJ...689.1327S} evolutionary models (left panel). In each panel, the data points are colored using the median value of $\log g$ derived using the same model as the effective temperature. The relationship between $T_\mathrm{eff}$ and spectral type for field objects described by \cite{2015ApJ...810..158F} is plotted as a dashed grey line and the dispersion is shown as a grey shadow.}
    \label{fig:teff-vs-spt}
\end{figure}

\begin{figure}[!t]
    \centering
    \includegraphics[width=\linewidth]{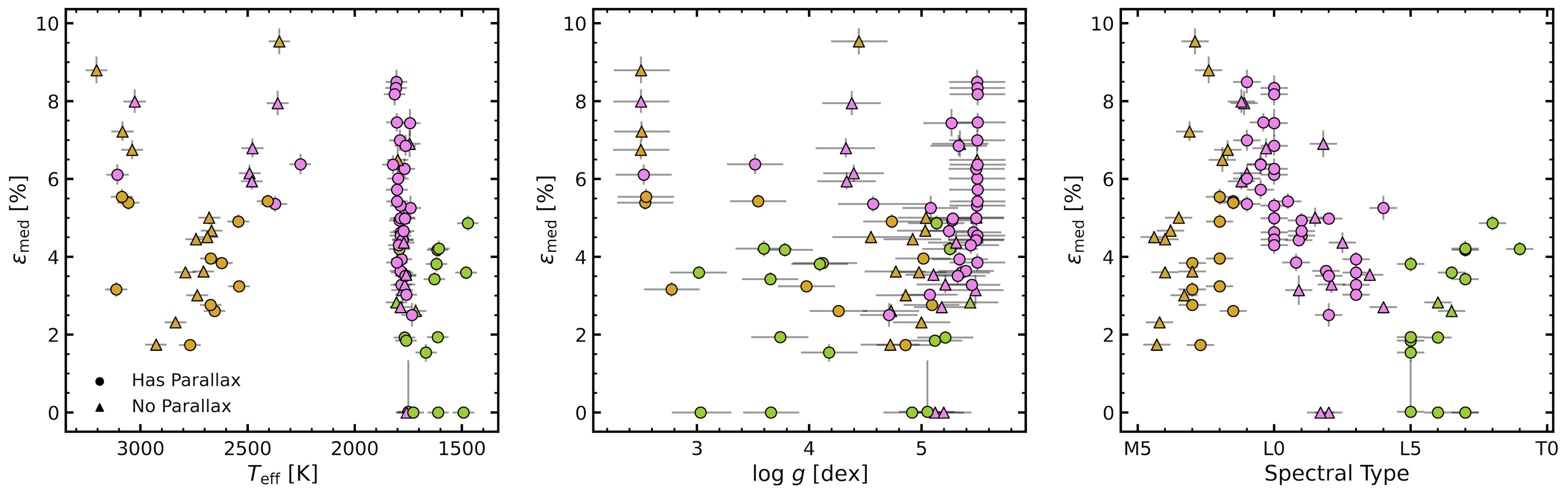}
    \caption{Plots of $\epsilon_\mathrm{med}$ versus $T_\mathrm{eff}$ (left panel), $\log g$ (middle panel), and spectral type (right panel). Orange points represent objects with spectral types M8.5 and earlier, pink spectral types M9 to L4, and green L4.5 and later.}
    \label{fig:epsilon-vs-params}
\end{figure}

\section{Evolutionary Model Analysis}
\label{sec:evolutionaryanalysis}

We derive the same set of physical parameters, as inferred from the spectral fitting, for the candidate moving group members using hot-start evolutionary models. Both the \cite{2008ApJ...689.1327S} hybrid model and \cite{2015A&A...577A..42B} model include cloud formation in their evolutionary sequences. However, some objects in our sample are too hot for the \cite{2008ApJ...689.1327S} hybrid model ($T_\mathrm{eff} \gtrsim 2500$ K) and others are too cool for the \cite{2015A&A...577A..42B} models ($T_\mathrm{eff} \lesssim 1500$ K). Therefore, we use either one of the cloudy evolutionary models or both when appropriate.

For each object, we begin by calculating the apparent bolometric magnitude using the analytic relationship between spectral type and the bolometric correction to $K_S$ photometry given by \cite{2015ApJ...810..158F} for young objects. We additionally calculate the uncertainty in the bolometric correction assuming the root mean square deviation from the relationship reported by \cite{2015ApJ...810..158F} and that the spectral type of each object has an uncertainty of $\pm0.5$ subtypes. 7 of our objects are missing 2MASS photometry but do have MKO photometry. For these objects, we synthesize offsets from the MKO $K$-band to the $K_S$-band using their observed spectra and the corresponding filter response curves \citep{2002A&A...386.1157G, 2003AJ....126.1090C}. We then use the object's distances to convert their apparent bolometric magnitudes into bolometric luminosities, $L_\mathrm{bol}$.

We additionally assume the object's age ($t$) falls within the distribution of ages associated with  its moving group (\rftabl{ymg}). To estimate the evolutionary-model parameters of our objects, we draw $\log L_\mathrm{bol}$ and $t$ posteriors using the Bayesian rejection sampling technique described in \cite{2017ApJS..231...15D}, where $10^6$ random samples are initially drawn from uniform distributions spanning the range of $\log L_\mathrm{bol}$ and $t$ values for both evolutionary models. For ages following uniform distributions, $\chi^2$ is computed for each sample as
\begin{equation}
    \chi^2 = \frac{\left(\log L_\mathrm{bol,model} - \log L_\mathrm{bol}\right)^2}{\sigma_{\log L_\mathrm{bol}}^2}
\end{equation}
while for ages following Gaussian distributions it is calculated as
\begin{equation}
    \chi^2 = \frac{\left(\log L_\mathrm{bol,model} - \log L_\mathrm{bol}\right)^2}{\sigma_{L_\mathrm{\log bol}}^2} + \frac{\left(t_\mathrm{model} - t\right)^2}{\sigma_t^2}.
\end{equation}
The probability associated with each sample is
\begin{equation}
    p = \exp\left[-\frac{1}{2}\left(\chi^2 - \chi_\mathrm{min}^2\right)\right],
\end{equation}
which has been normalized by the sample with the lowest $\chi^2$. We draw random variates ($u$) from a uniform distribution ranging from 0 and 1 and reject samples where $p<u$, along with any samples that fall outside of the evolutionary model parameter space. We repeat this process twice more, successively narrowing the trial bolometric luminosity and age distribution to improve sampling efficiency.

We then linearly interpolate the evolutionary models and calculate the $\log T_\mathrm{eff}$, $\log g$, $\log R$, and $\log M$ associated with each sample in our final $\log L_\mathrm{bol}$ and $\log t$ posteriors. The median effective temperature, surface gravity, radius, and mass calculated for each object using the \cite{2008ApJ...689.1327S} and/or \cite{2015A&A...577A..42B} evolutionary models are given with 68\% confidence intervals in \rftabl{derived_evo}. As seen in \rffigl{evovevo}, the two models are nearly always consistent with each other; however, the \cite{2015A&A...577A..42B} model returns median $T_\mathrm{eff}$ typically higher ($\mysim25$ K) and median radii and masses typically lower ($\mysim0.04 \ R_\mathrm{Jup}$ and $\mysim2 \ M_\mathrm{Jup}$, respectively), than those returned by the \cite{2008ApJ...689.1327S} hybrid model.

\section{Assessing Atmospheric Model Performance}
\label{sec:benchmarking}

\subsection{Systematic Behavior in Fitted and Derived Parameters}
\label{sec:parameters}

Predictions of model atmospheres have well-recognized systematic uncertainties that can bias the derived physical parameters (e.g., \citealt{2008ApJ...678.1372C, 2009ApJ...702..154S, 2017ApJ...842..118L}). On the other hand, masses and radii predicted by evolutionary models often match those found for ultracool dwarfs with dynamical masses (e.g., \citealt{2017ApJS..231...15D, 2019AJ....158..140B, 2021AJ....162..301B}) and transiting brown dwarfs with directly measured radii (e.g., \citealt{2016ApJ...822L...6M, 2020AJ....160...53C, 2020NatAs...4..650T}). Throughout this work, we assume our evolutionary model parameters to be more robust, and use them to test the systematic uncertainties in the predictions of the atmospheric models.

We compare our spectroscopically inferred physical parameters against those derived using the \cite{2015A&A...577A..42B} evolutionary model in \rffigl{btsettl-vs-baraffe} and against those derived using the \cite{2008ApJ...689.1327S} hybrid evolutionary model in \rffigl{btsettl-vs-sm08}. BT-Settl struggles to reproduce most of the evolutionary $T_\mathrm{eff}$ below $\sim$2500 K, instead returning temperatures clustered near 1800 K.

Plotting $T_\mathrm{eff}$ versus spectral type (see \rffigl{teff-vs-spt}) reveals that our evolutionary-based temperatures generally follow the analytic relationship for field objects described by \cite{2015ApJ...810..158F}, as expected since this relationship is established using evolutionary sequences. Further, our bolometric corrections were calculated as a function of spectral type using a polynomial relation also given by \cite{2015ApJ...810..158F} for young objects. However, likely because our sample consists mostly of young objects, evolutionary temperatures for late M-dwarfs appear slightly hotter than the \cite{2015ApJ...810..158F} relation while L dwarfs are overall slightly cooler. In contrast, the BT-Settl temperatures cluster near 1800 K for objects at the M-L spectral type boundary (roughly from M8 to L4). BT-Settl additionally returns $\log g$ posteriors pressed against the upper limit of the model grid (5.5 dex) for nearly all of these objects. These high spectroscopically inferred surface gravities imply very large masses of 150-1400~$M_\mathrm{Jup}$ while we anticipate masses $\lesssim70~M_\mathrm{Jup}$ for our sample. The evolutionary models appear to return a more realistic distribution of surface gravities, although we note that both \cite{2008ApJ...689.1327S} and \cite{2015ApJ...810..158F} tend to return higher surface gravities for earlier objects.

\begin{figure}
    \centering
    \includegraphics[width=\linewidth]{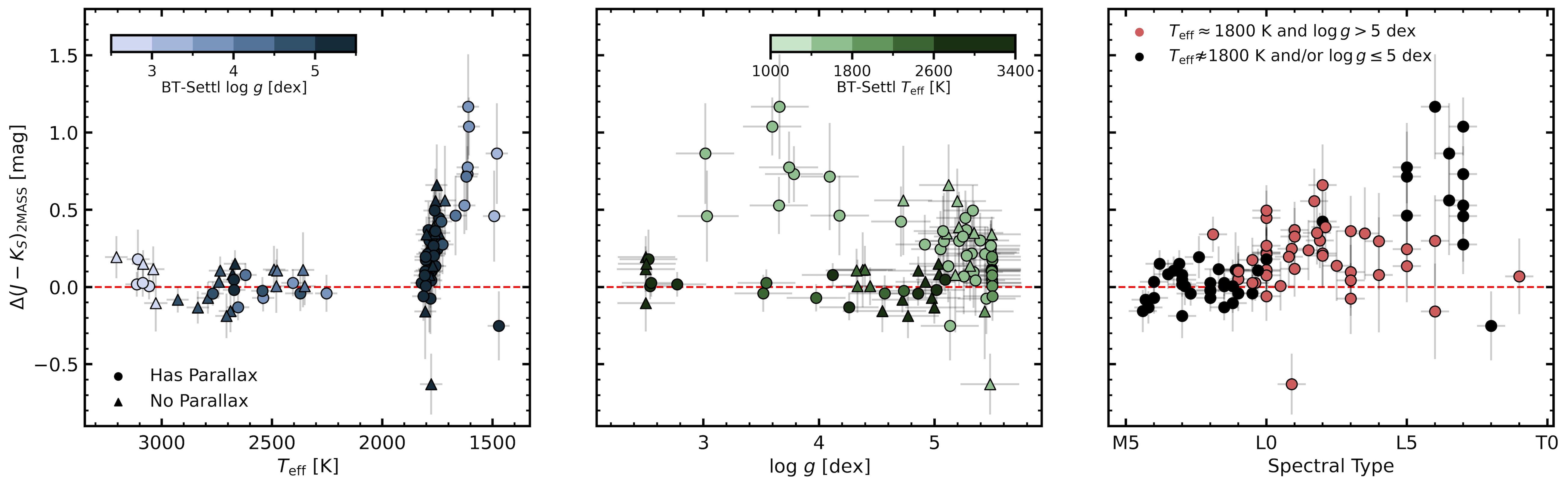}
    \caption{The color anomaly, $\Delta(J-K_S)_\mathrm{2MASS}$, for each object in our sample with 2MASS photometry. Positive values indicate that an object is unusually red while negative values indicate that an object is unusually blue. In the left panel, the anomaly is plotted against the $T_\mathrm{eff}$ derived using BT-Settl and data points are colored according to the $\log g$ derived from BT-Settl. In the middle panel, the anomaly is plotted against $\log g$ and colored according to $T_\mathrm{eff}$. In the left panel, $\Delta(J-K_S)_\mathrm{2MASS}$ is plotted against spectral type; objects with derived $T_\mathrm{eff}$ clustered around 1800 K and elevated $\log g$ ($>5$ dex) are indicated by red data points.}
    \label{fig:jkanomalies}
\end{figure}

\begin{figure}
    \centering
    \includegraphics[width=0.9\linewidth]{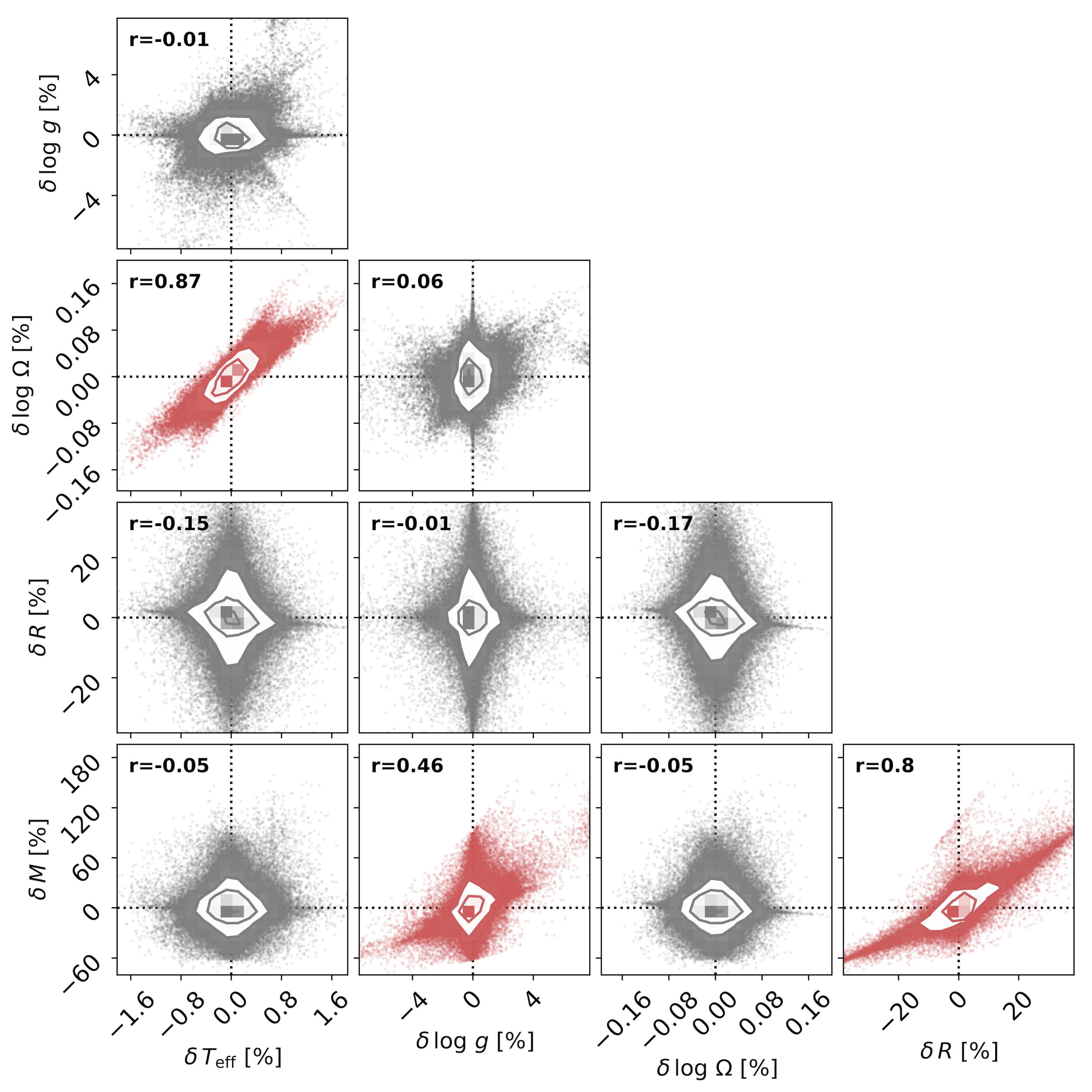}
    \caption{Stacked posteriors for the five physical parameters determined from atmospheric modeling for all 90 objects in our sample. The fractional deviation from the median of the posteriors was calculated for each sample, leaving chains that represent the relative variation in each posterior, denoted as $\delta T_\mathrm{eff}$, $\delta\log g$, $\delta\log\Omega$, $\delta R$, and $\delta M$. We combine the modified chains to generate the stacked posteriors and calculate the Spearman rank correlation coefficient for each pair of parameters (given in the upper left corner of each panel). The p-values associated with each coefficient are below $10^{-34}$. Most parameters have weak correlations with coefficients below 0.25 (indicated by grey). Three pairs of parameters have strong correlations with coefficients of 0.4 or above (marked by red), including $T_\mathrm{eff}$/$\log\Omega$, $M$/$R$, and $M$/$\log g$. Each of these strong correlations are expected.}
    \label{fig:stackedposteriors}
\end{figure}

BT-Settl also returns elevated $T_\mathrm{eff}$ above $\mysim$3000 K for 7 objects with spectral types ranging roughly from M8 to L0, although several other objects in our sample within this spectral type range have spectroscopically inferred temperatures consistent with evolutionary models and the analytic relationship from \cite{2015ApJ...810..158F}. These objects have very low surface gravities (below 3 dex) from the fitted spectra, which imply non-physical masses between 0.02 $M_\mathrm{Jup}$ and 3 $M_\mathrm{Jup}$.

\begin{figure}
    \centering
    \includegraphics[width=\linewidth]{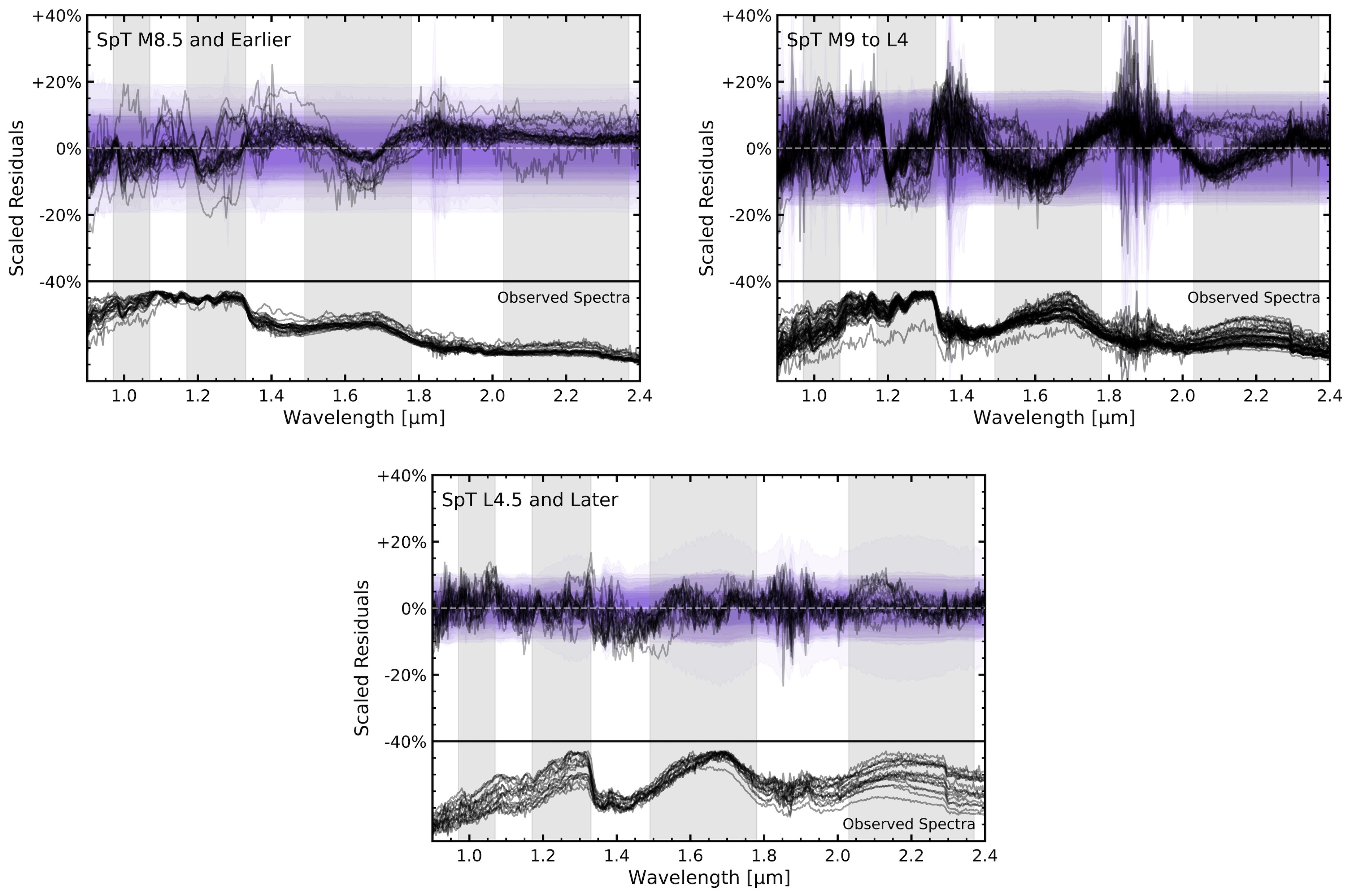}
    \caption{Stacked spectral-fitting residuals (using the median BT-Settl parameters, normalized by the peak observed flux, and shown as black lines) for objects with spectral types M8.5 and earlier (top left panel), objects with spectral types ranging from M9 to L4 (top right panel), and those with spectral types L4.5 and later (bottom panel). The $1\sigma$ and $2\sigma$ dispersions (corresponding to the flux uncertainties and $\epsilon_\mathrm{med}$ terms added in quadrature) have been plotted for each object as purple shadows. The normalized spectra for objects in each spectral type bin have been plotted at the bottom of all three panels as a reference. The wavelength intervals corresponding to the $Y$, $J$, $H$, and $K$ bands are marked by grey regions (the boundaries for each band are given in Equations \ref{eq:qY}, \ref{eq:qJ}, \ref{eq:qH}, and \ref{eq:qK}.}
    \label{fig:stackedresiduals}
\end{figure}

\begin{figure}
    \centering
    \includegraphics[width=\linewidth]{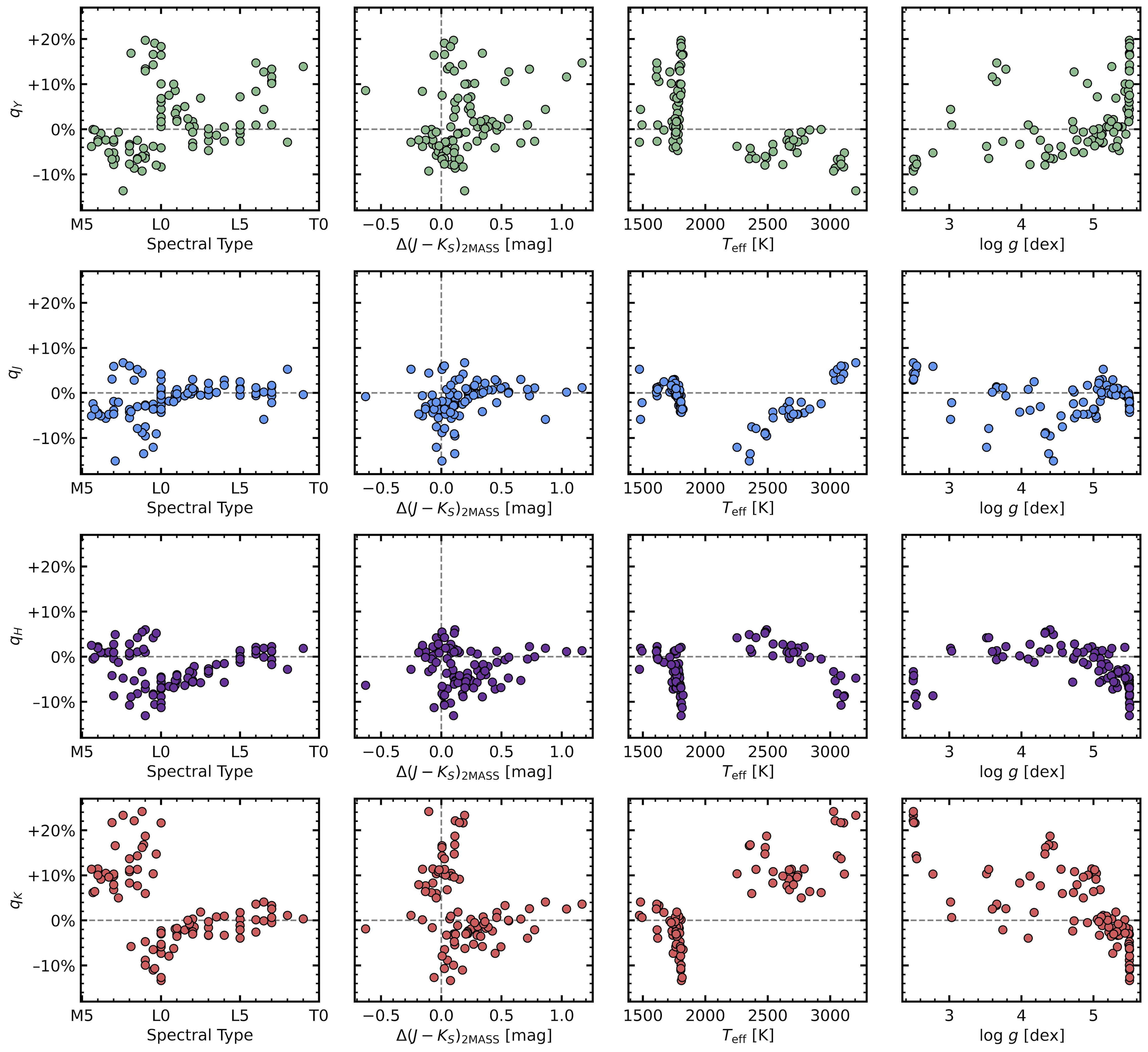}
    \caption{$q_Y$ (top row), $q_J$ (second row), $q_H$ (third row), and $q_K$ (bottom row) corresponding to the observed spectra and median BT-Settl models for all 90 objects in our sample as a function of spectral type (left column), color anomaly (second column), and $T_\mathrm{eff}$ (third column) and $\log g$ (fourth column) derived using BT-Settl. The $q$ values (Equations \ref{eq:q}-\ref{eq:qK}) quantify the residuals between our data and the fitted spectra in the $Y$, $J$, $H$, and $K$ bands. Negative values of $q$ indicate that the model overpredicts the flux in a band whereas positive values indicate that the model underpredicts the flux. Larger values of $q$ correspond to greater discrepancies between our data and the models.}
    \label{fig:residuals_quantified}
\end{figure}

Larger jitter terms are fit to objects poorly described by model atmospheres, capturing structure in the residuals. When plotting $\epsilon_\mathrm{med}$ (the jitter normalized to the median flux of the observed spectrum) versus $T_\mathrm{eff}$, $\log g$, and spectral type in \rffigl{epsilon-vs-params}, a clear peak emerges near the M-L transition boundary. Visual inspection of Figures \ref{fig:allspectra_nopleiades}-\ref{fig:allspectra_pleiades} confirm that BT-Settl most struggles to describe these objects. To investigate whether this systematic behavior can be attributed to anomalous infrared colors, we calculate $\Delta(J-K_S)_\mathrm{2MASS}$, defined as the difference between the observed color, $(J-K_S)_\mathrm{2MASS}$, and the typical color for a field object with a given spectral type \citep{2018ApJS..234....1B}. While it is conventionally expected that the young objects in our sample will appear redder than field objects, \cite{2016ApJ...821..120A} note that ABDMG objects, some of the oldest in this work, have similar colors to field objects. \cite{2016ApJ...833...96L} and \cite{2016ApJS..225...10F} suggest that redder colors are not strictly indicative of younger ages, while \cite{2017ApJ...842...78V} and \cite{2023ApJ...954L...6S} find that factors such as viewing geometries can influence object colors. Regardless, we find that typical field colors are useful in gauging the color anomaly for both field and YMG objects. For objects that lack 2MASS photometry, we calculate $\Delta(J-K_S)_\mathrm{2MASS}$ using magnitudes synthesized from the observed spectra. Positive values of $\Delta(J-K_S)_\mathrm{2MASS}$ indicate that an object is relatively red for its spectral type while negative values indicate that it is relatively blue. \rffigl{jkanomalies} shows that many of our objects located near the M-L transition boundary have $\Delta(J-K)_\mathrm{2MASS}$ consistent with 0, including those with $T_\mathrm{eff}$ clustered above 3000 K and very low surface gravities. However, a large number of the objects clustered near $T_\mathrm{eff}$ of 1800 K and that have high surface gravities are abnormally red. This suggests that while BT-Settl struggles to model all objects near the M-L transition boundary, the systematic behavior in the derived $T_\mathrm{eff}$ and $\log g$ are ultimately determined by the color of an object.

To search for additional systematic behavior such as correlations between physical parameters, we stack the posteriors derived using atmospheric models for all our objects following a procedure similar to that in \cite{2021ApJ...921...95Z}. First, we calculate the deviation from the median of the posteriors for each object's MCMC chain, leaving values that represent the relative variation in each posterior, which we denote as $\delta T_\mathrm{eff}$, $\delta\log g$, $\delta\log\Omega$, $\delta R$, and $\delta M$. We then concatenate the modified chains to produce the stacked posteriors, which are shown in \rffigl{stackedposteriors}. Using these stacked posteriors, we calculate the Spearman rank correlation coefficient for each pair of parameters. Three strong correlations are found. First, $T_\mathrm{eff}$ and $\log\Omega$ have a correlation coefficient of 0.87; at face value, this is unexpected given that effective temperature and solid angle are negatively correlated via the Stefan-Boltzmann law. However, as discussed in \rfsecl{atmosphericanalysis}, the BT-Settl models are not strictly monotonic for many objects in our sample and the variation in $T_\mathrm{eff}$ more likely represents changes in the morphology of model spectra than in luminosity. $M$ and $R$ have a coefficient of 0.8 while $M$ and $\log g$ have a coefficient of 0.46; these are expected, given that the mass is derived from the radius and surface gravity and is more strongly dependent on radius.

\begin{figure}
    \centering
    \includegraphics[width=\linewidth]{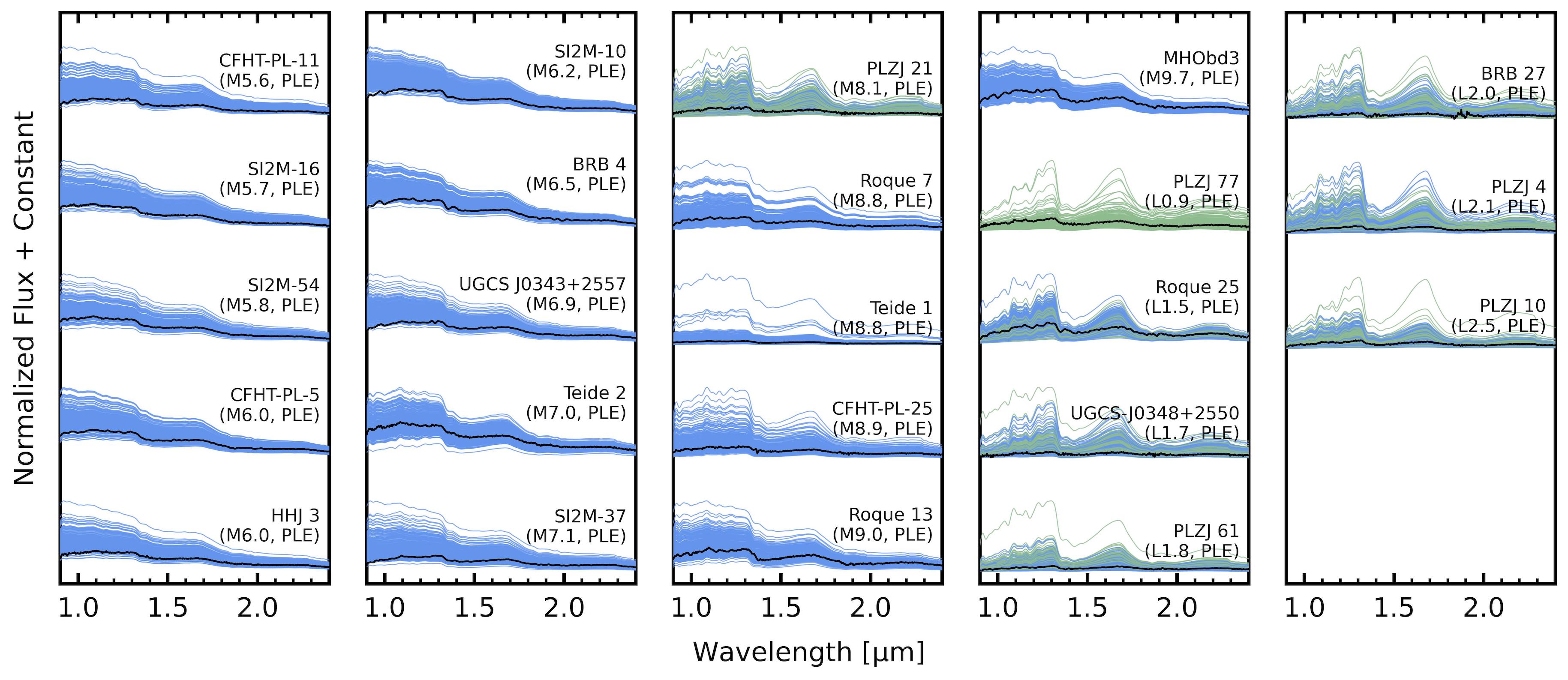}
    \caption{Observed spectra as black lines plotted against 500 BT-Settl models drawn from the evolutionary model posteriors for each object in our sample located in the Pleiades. Blue lines correspond to the \cite{2015A&A...577A..42B} model posterior and green the \cite{2008ApJ...689.1327S} hybrid model posterior. For each object, the models were first scaled using the radius and distance; the observed spectrum and models were then normalized using the same constant. The models show much more variation than the ones plotted in \rffigl{allspectra_pleiades} because the posteriors returned by the evolutionary models are much broader than those directly output by forward modeling. BT-Settl models corresponding to evolutionary parameters consistently overpredict the observed flux.}
    \label{fig:allevospectra_pleiades}
\end{figure}

\begin{figure}
    \centering
    \includegraphics[width=\linewidth]{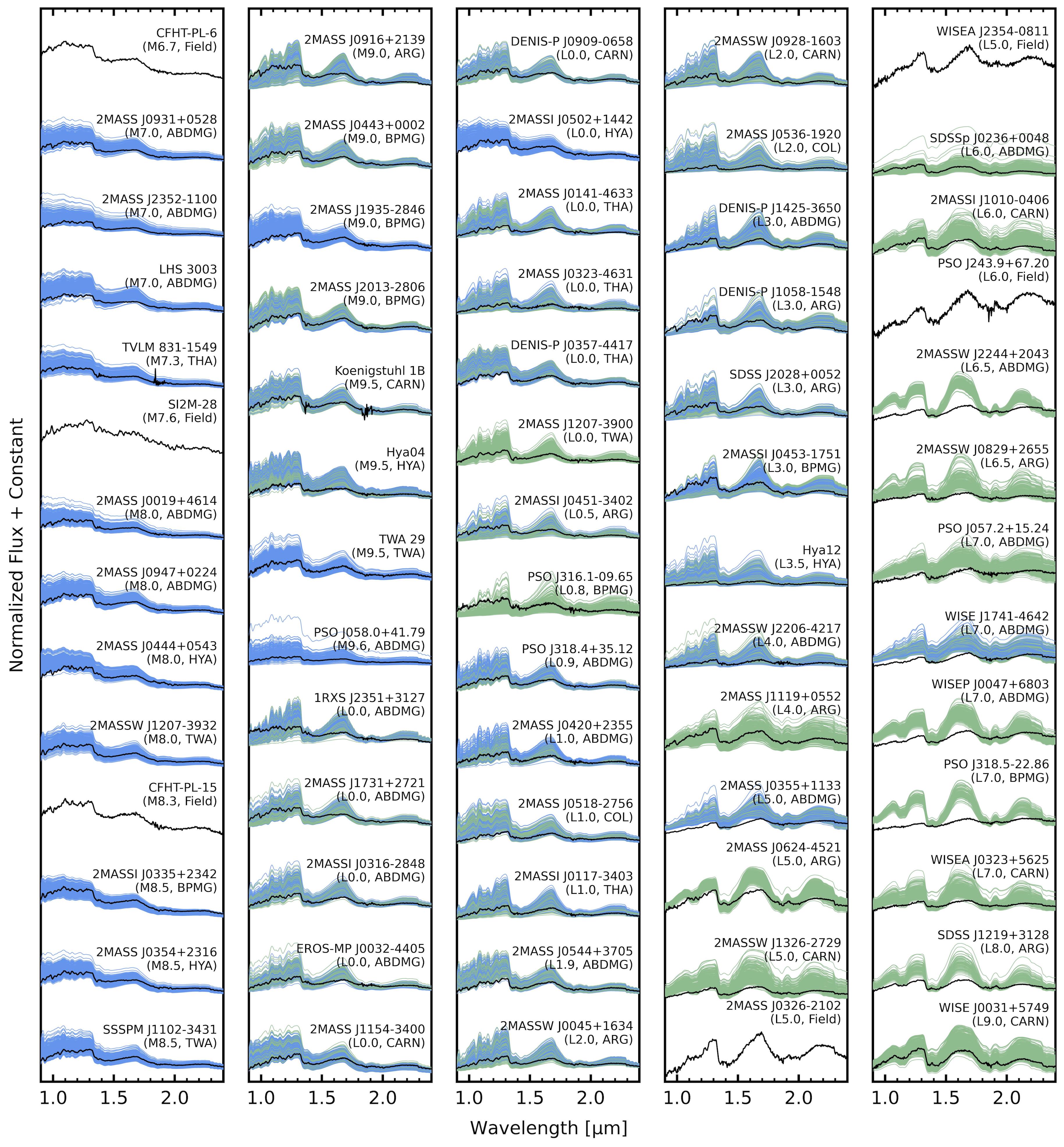}
    \caption{Observed spectra as black lines plotted against 500 BT-Settl models drawn from the evolutionary model posteriors for each object in our sample located outside of the Pleiades, with blue corresponding to the \cite{2015A&A...577A..42B} model posterior and green the \cite{2008ApJ...689.1327S} hybrid model posterior. For each object, the models were first scaled using the radius and distance; the observed spectrum and models were then normalized using the same constant. No models are shown for the six field objects which have unknown ages and cannot be placed on evolutionary sequences. The models show much more variation than those in \rffigl{allspectra_nopleiades} because the posteriors returned by the evolutionary models are much broader than those directly output by forward modeling. BT-Settl models corresponding to evolutionary parameters consistently overpredict the observed flux.}
    \label{fig:allevospectra_nopleiades}
\end{figure}

\begin{figure}
    \centering
    \includegraphics[width=\linewidth]{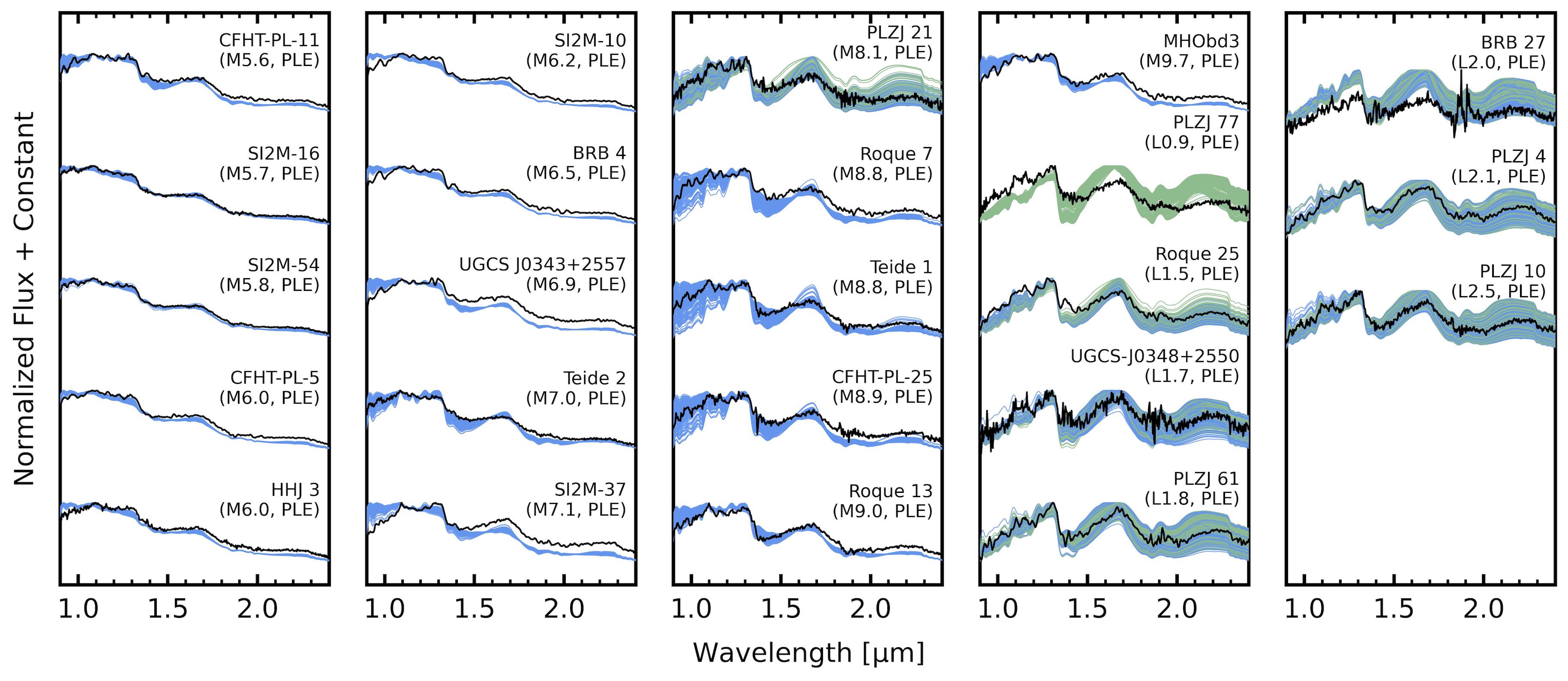}
    \caption{The same as \rffigl{allevospectra_pleiades}, except each model and observed spectrum has been normalized using its maximum flux. This eases comparisons between the morphology of model and observed spectra. Evolutionary parameters return model spectra than trend bluer than the observed spectrum for objects near the M-L transition boundary. When evolutionary parameters return model spectra with colors similar to those of the observed spectrum, the $H$-band region is often flatter in the observed spectrum than in the model spectra. Overall, the evolutionary parameters appear to return poorly-fitting model spectra.}
    \label{fig:allevospectra_rescaled_pleiades}
\end{figure}

\begin{figure}
    \centering
    \includegraphics[width=\linewidth]{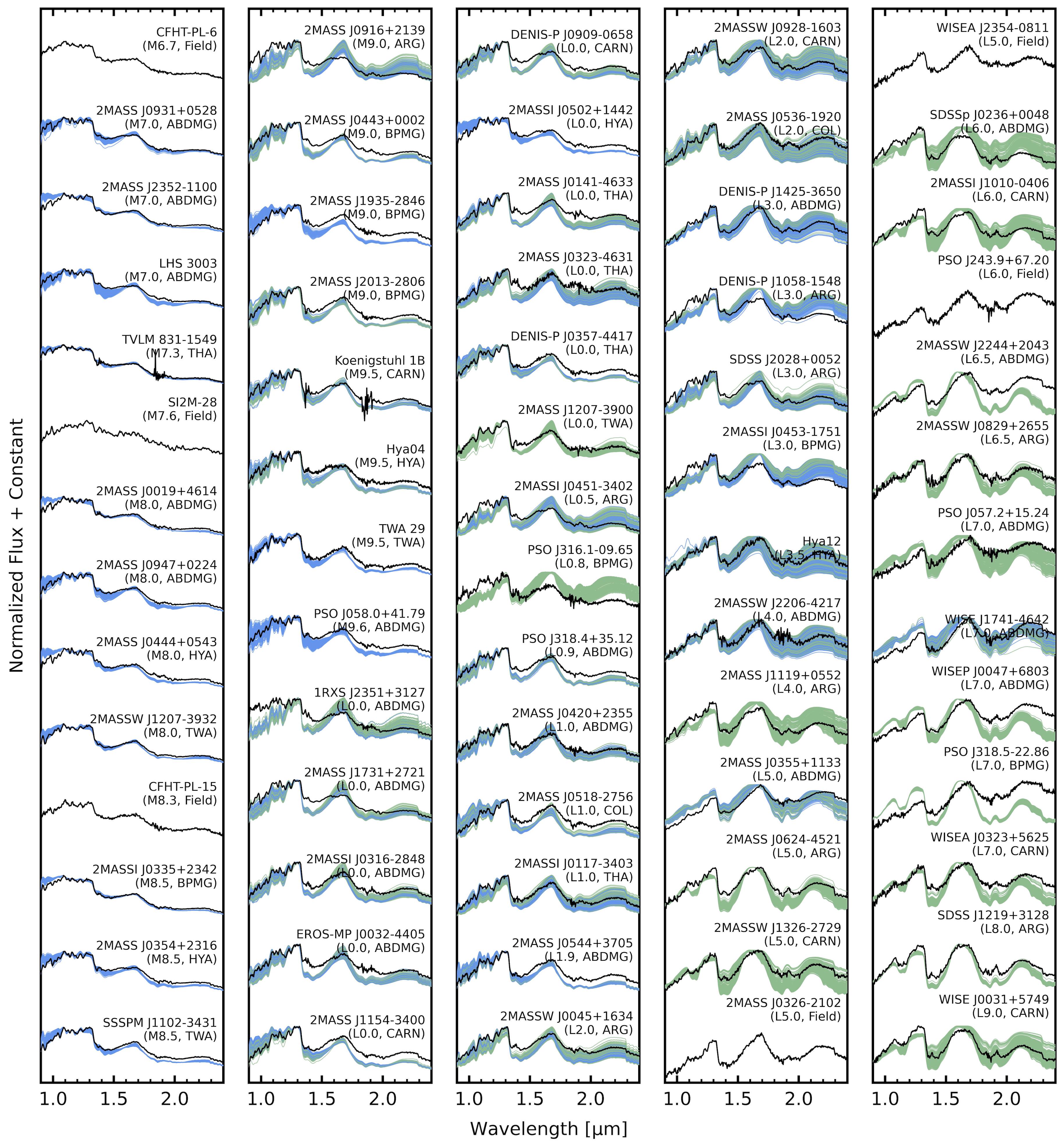}
    \caption{The same as \rffigl{allevospectra_nopleiades}, except each model and observed spectrum has been normalized using its maximum flux. This eases comparisons between the morphology of model and observed spectra. Evolutionary parameters return model spectra than trend bluer than the observed spectrum for objects near the M-L transition boundary. When evolutionary parameters return model spectra with colors similar to those of the observed spectrum, the $H$-band region is often flatter in the observed spectrum than in the model spectra. Overall, the evolutionary parameters appear to return poorly-fitting model spectra.}
    \label{fig:allevospectra_rescaled_nopleiades}
\end{figure}

\subsection{Systematic Behavior in Model Spectra and Residuals}
\label{sec:residuals}

As discussed in \rfsecl{parameters}, the BT-Settl models consistently fit objects near the M-L spectral type transition boundary poorly; to investigate whether there is systematic structure to the fitted spectra and residuals between our models and data, we plot the residuals (data$-$model) for each object (normalized using the peak observed flux) against each other in \rffigl{stackedresiduals}. Here, we see that the fitted BT-Settl models tend to overpredict the peak fluxes in the $J$ and $H$ bands for objects with spectral types ranging from M9 to L4, along with several earlier spectral type objects. The residuals show a wider range of behavior in the $K$-band; BT-Settl overpredicts the flux at these wavelengths for many objects, yet the residuals are relatively small or show little structure. Alternatively, the model atmospheres underpredict the flux in the $K$-band for several other objects.

The peak fluxes in the $J$, $H$, and $K$ bands emerge from the deep layers of an object's photosphere, where the atmosphere is hot and luminous. The fact that BT-Settl overpredicts the $J$ and $H$-band fluxes for objects near the M-L transition boundary suggests that these models do not include sufficient dust opacity to extinct the light from these deep layers. This interpretation is consistent with the shallower residuals observed in the $K$-band, where the dust extinction effect is intrinsically smaller than in the $J$ and $H$ bands. 

To better quantify the residuals in each band, we calculate the quantities $q_Y$, $q_J$, $q_H$, and $q_K$ using definitions similar to those defined in \cite{2021ApJ...921...95Z}:
\begin{align}
    \label{eq:q}
    q(\lambda\lambda) = 1 - \frac{\int_{\lambda\lambda}f_\mathrm{model,\lambda}\,d\lambda}{\int_{\lambda\lambda}f_\mathrm{obs,\lambda}\,d\lambda}, \\[1ex] \label{eq:qY} q_Y = q([0.97 \ \mathrm{\mu m}, \ 1.07 \ \mathrm{\mu m}]), \\ \label{eq:qJ} q_J = q([1.17 \ \mathrm{\mu m}, \ 1.33 \ \mathrm{\mu m}]), \\ \label{eq:qH} q_H = q([1.49 \ \mathrm{\mu m}, \ 1.78 \ \mathrm{\mu m}]), \\ \label{eq:qK} q_K = q([2.03 \ \mathrm{\mu m}, \ 2.37 \ \mathrm{\mu m}]).
\end{align}
Unlike \cite{2021ApJ...921...95Z}, the $\lambda\lambda$ used in this work do correspond to standard definitions of the $Y$, $J$, $H$, and $K$ filters \citep{2002PASP..114..169S, 2002PASP..114..180T, 2005PASP..117..421T}. Additionally, $f_\mathrm{model,\lambda}$ and $f_\mathrm{obs,\lambda}$ represent the fitted model (we use the model calculated from the median parameter values) and observed spectrum, respectively. Positive values of $q$ indicate that the model underpredicts the flux in a given wavelength range while negative values indicate that the model overpredicts the flux. Larger absolute values of $q$ correspond to greater discrepancies between the models and data. These quantities are plotted versus spectral type, color anomaly ($\Delta(J-K_S)_\mathrm{2MASS}$) and physical parameters derived using BT-Settl, including $T_\mathrm{eff}$ and $\log g$, in \rffigl{residuals_quantified}.

\begin{figure}
    \centering
    \includegraphics[width=\linewidth]{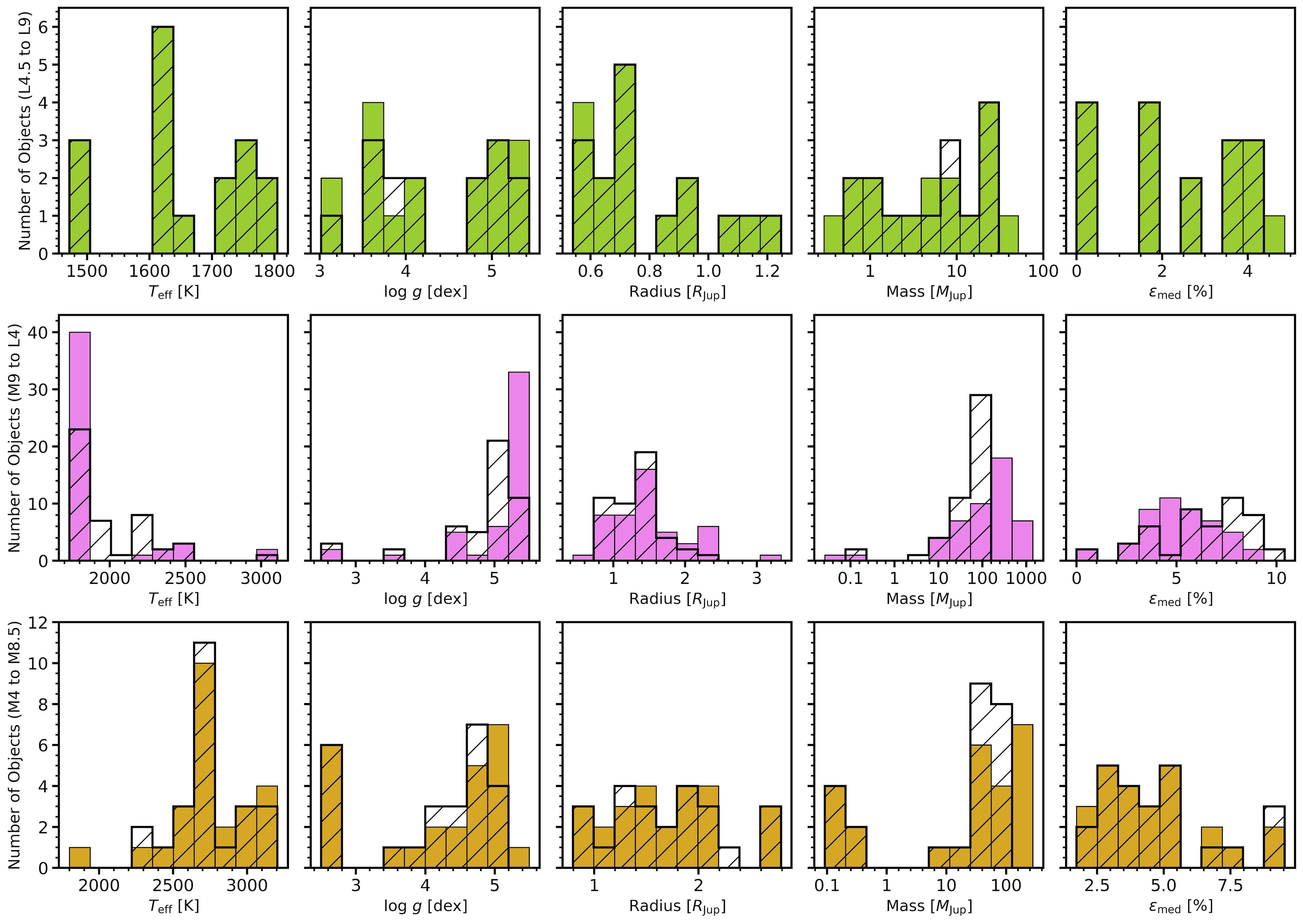}
    \caption{Histograms of parameters returned by BT-Settl without a mass prior (filled, colored bars) and with a mass prior (hatched, colorless bars). The top row corresponds to objects with spectral types L4.5 and later, the middle row spectral types M9 to L4, and the bottom row M8.5 and earlier.}
    \label{fig:prior-vs-noprior-hist}
\end{figure}

The residual structure revealed by $q$ for objects near the M-L transition boundary varies widely, despite objects having the same or similar spectral types. Objects with $T_\mathrm{eff}$ clustered near 1800 K and high $\log g$ tend to have underpredicted fluxes by BT-Settl in the $Y$-band ($q_Y$ up to 20\%), moderately overpredicted fluxes in the $J$-band ($q_J$ down to -5\%), and more dramatically overpredicted fluxes in the $H$ and $K$ bands ($q_H$ and $q_K$ down to -15\%). The objects with $T_\mathrm{eff}$ clustered above 3000 K and low $\log g$ have overpredicted fluxes by BT-Settl in the $Y$ and $H$ bands ($q_Y$ and $q_H$ down to -10\%) but fluxes moderately underpredicted in the $J$-band ($q_J$ near 5\%) and fluxes greatly underpredicted in the $K$-band ($q_K$ as high as 25\%). Interestingly, objects near the M-L boundary show a sharp divide in their $Y$ and $K$-band residuals; earlier objects near the boundary are overpredicted in the $Y$-band and underpredicted in the $K$-band while the opposite is true for later objects.

To further search for systematic behaviors in BT-Settl, we plot model spectra corresponding to the evolutionary model posteriors against our data in \rffigl{allevospectra_pleiades} and \rffigl{allevospectra_nopleiades}. We scale the model spectra using the evolutionary radii posteriors and object distances. Because the evolutionary sequences return broader posteriors than those taken directly from spectral fitting, we see more variation in different models plotted for the same object. However, our observed spectra tend to be dimmer than the majority of the corresponding models at all wavelengths, suggesting that spectral fitting with BT-Settl has a tendency to underestimate either $T_\mathrm{eff}$ or $R$. To more directly compare spectral morphologies, we also plot model spectra corresponding to the evolutionary model posteriors against our data in \rffigl{allevospectra_rescaled_pleiades} and \rffigl{allevospectra_rescaled_nopleiades}, but have normalized each individual model and observed spectrum using its maximum flux. We see that evolutionary parameters return model spectra bluer than the observed spectrum for many objects near the M-L transition boundary. When evolutionary parameters return model spectra with colors similar to those of the observed spectrum, the model spectra consistently overpredict the $H$-band flux for objects near the M-L transition boundary, featuring a triangular-shaped continuum often indicative of low gravities \citep{2013ApJ...772...79A}. Given that our evolutionary models return surface gravities between 4.0 and 5.5 dex and that many of our spectra have relatively flat continua within the $H$-band when compared to the model spectra, it appears that BT-Settl fails to replicate features indicative of moderate and high surface gravities. We also note that the models corresponding to the evolutionary posteriors match our data very poorly for several objects with spectral types ranging from L6.5 to L8. Yet, the models fit using forward modeling match the data very closely for these objects (see \rffigl{allspectra_nopleiades}). This indicates that while some BT-Settl spectra can describe our data quite well, they return biased temperatures and gravities, although this discrepancy may be explained if some of our candidates are not true moving group members with young ages.

\begin{figure}
    \centering
    \includegraphics[width=\linewidth]{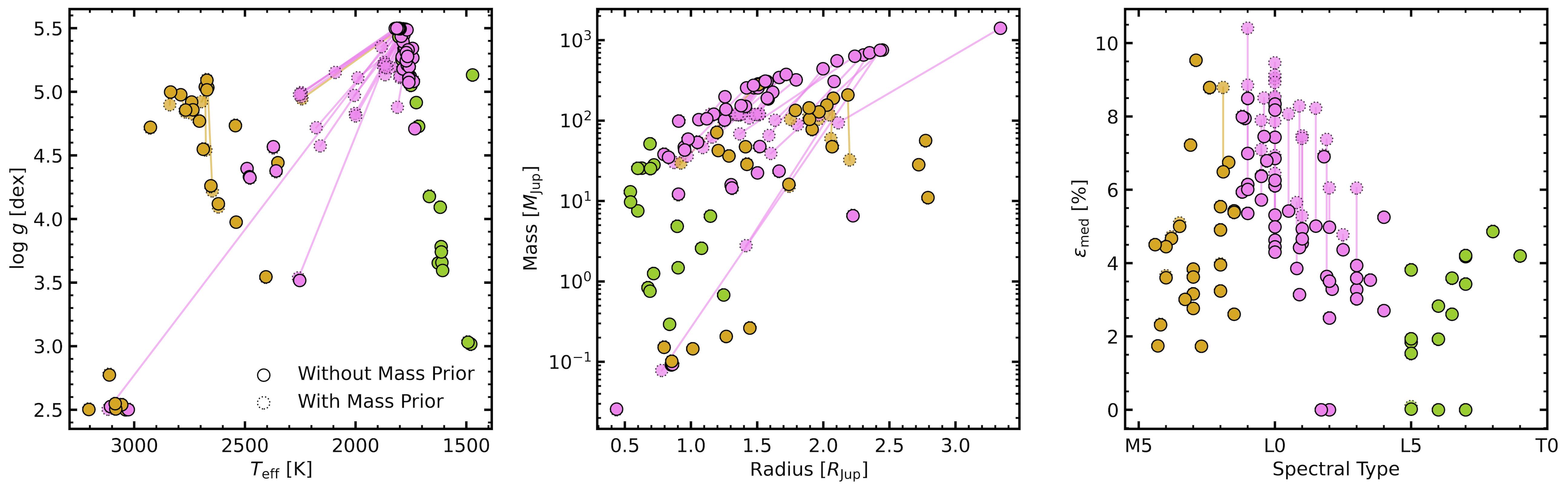}
    \caption{Points with solid edges represent the median parameters returned by BT-Settl without a mass prior while transparent points with dashed edges represent the median parameters returned with a mass prior. Lines connect points corresponding to the same object. Orange points represent objects with spectral types M8.5 and earlier, pink spectral types M9 to L4, and green L4.5 and later.}
    \label{fig:prior-vs-noprior-scatter}
\end{figure}

\begin{figure}
    \centering
    \includegraphics[width=\linewidth]{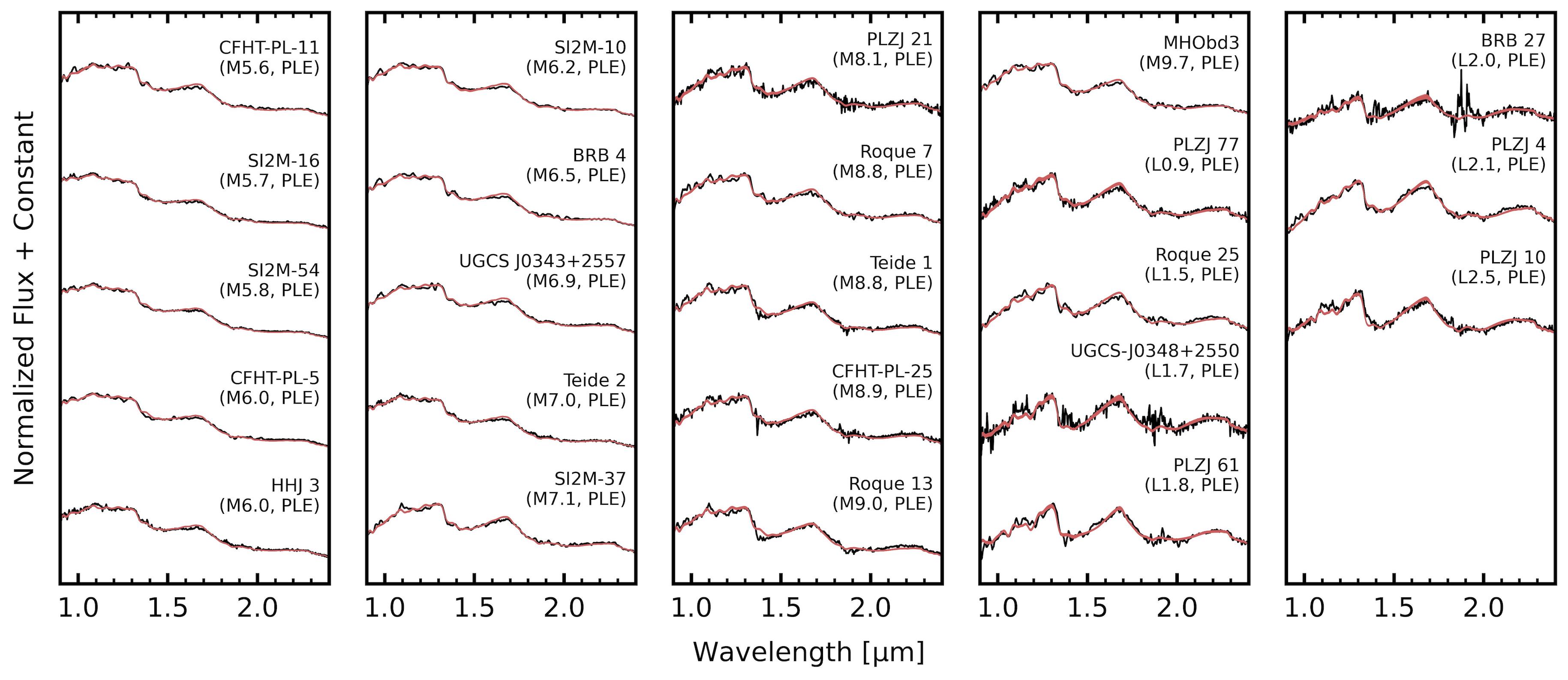}
    \includegraphics[width=\linewidth]{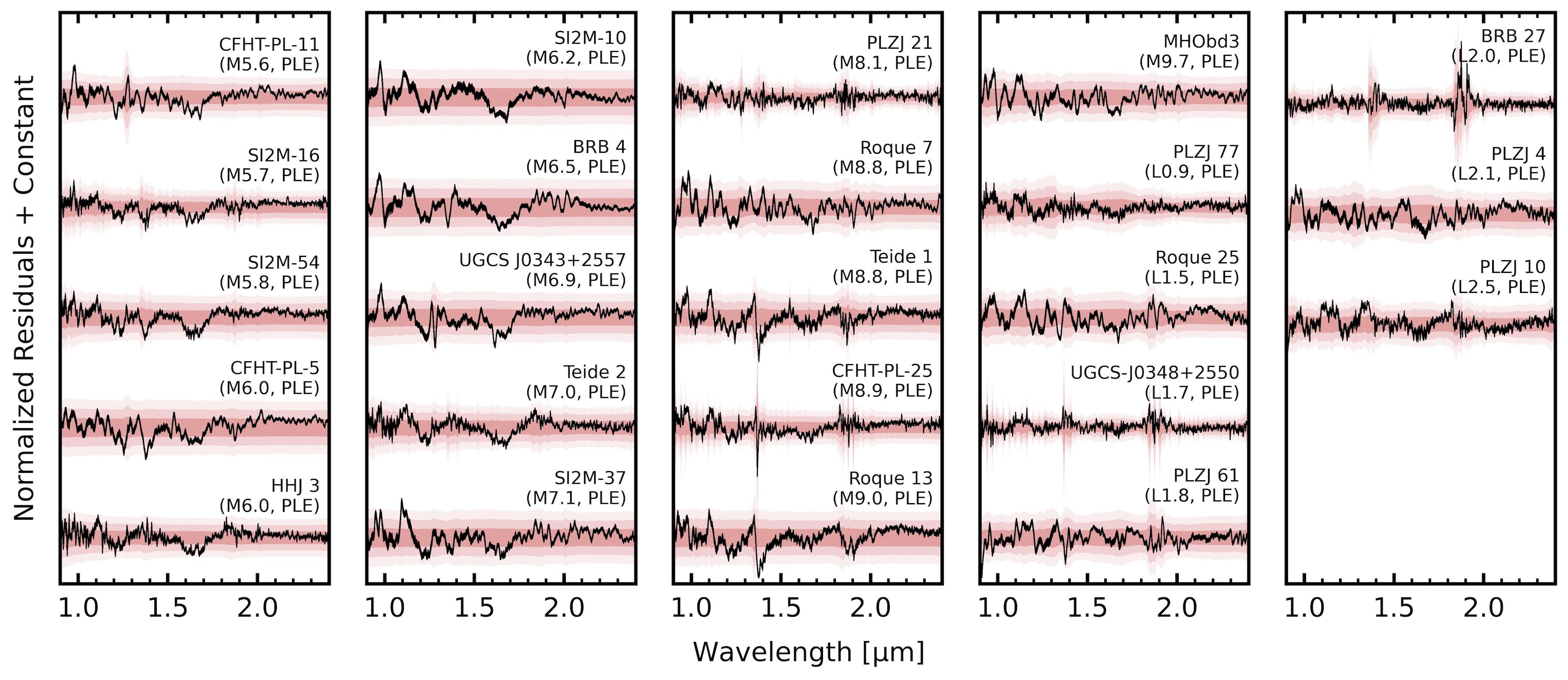}
    \caption{The top figure shows the observed spectra as black lines plotted against 250 BT-Settl models including interstellar extinction (red lines) drawn from the posteriors for each object located in the Pleiades. These posteriors come directly from the formal spectral fitting and do not include systematic uncertainties. The bottom figure shows the corresponding residuals as black lines along with the 1$\sigma$, 2$\sigma$, and  3$\sigma$ dispersions (the measurement uncertainties and $\epsilon_\mathrm{med}$ added in quadrature) as red shadows.}
    \label{fig:allspectra_pleiades_ext}
\end{figure}

\begin{figure}
    \centering
    \includegraphics[width=\linewidth]{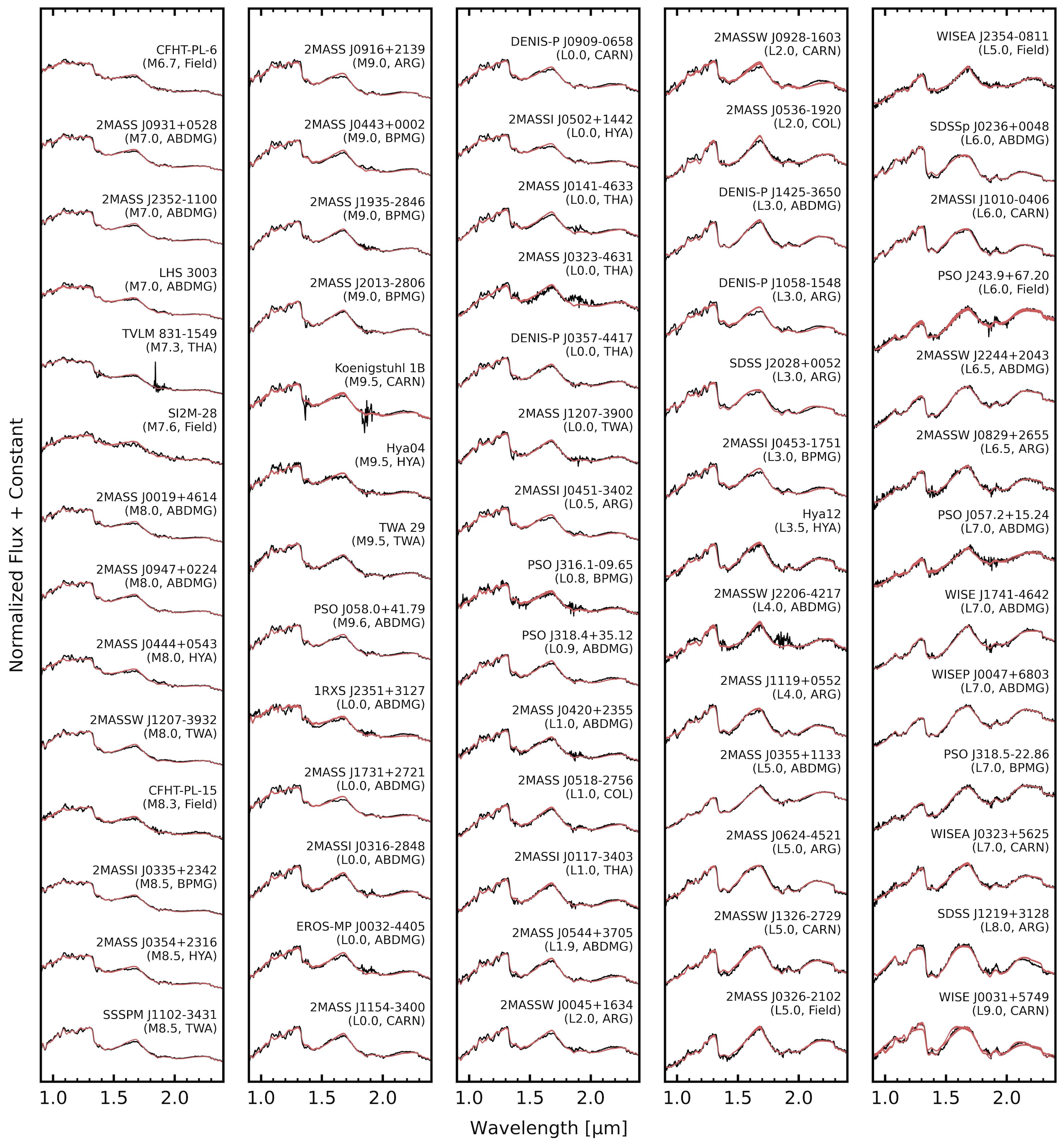}
    \caption{Observed spectra shown as black lines plotted alongside 250 BT-Settl models including interstellar extinction (red lines) drawn from the posteriors for each YMG and Hyades object in our sample. These posteriors come directly from the formal spectral fitting and do not include systematic uncertainties.}
    \label{fig:allspectra_nopleiades_ext}
\end{figure}

\begin{figure}
    \centering
    \includegraphics[width=\linewidth]{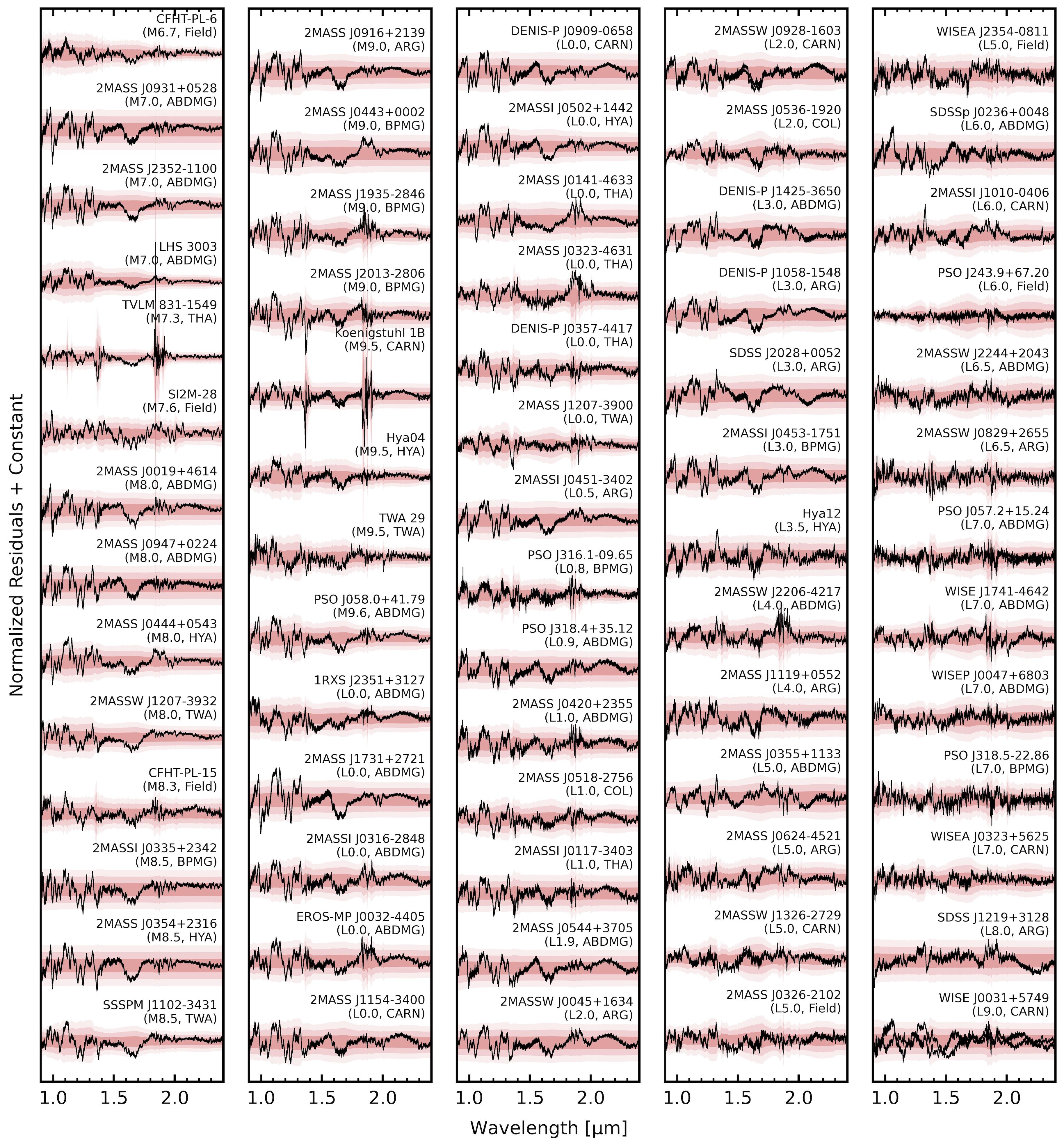}
    \caption{The residuals (data$-$model) between the observed spectra and 250 BT-Settl models including interstellar extinction drawn from the posteriors for each YMG and Hyades object in our sample are shown in black. These posteriors come directly from the formal spectral fitting and do not include systematic uncertainties. The 1$\sigma$, 2$\sigma$, and  3$\sigma$ dispersions (corresponding to the measurement uncertainties and $\epsilon_\mathrm{med}$ jitter) added in quadrature are shown as red shadows.}
    \label{fig:allresiduals_nopleiades_ext}
\end{figure}

\subsection{Including a Mass Prior}
\label{sec:prior}

Dynamical masses show L and late M dwarfs rarely have masses above 120 $M_\mathrm{Jup}$ \citep{2017ApJS..231...15D, 2019ApJ...871...63M}. We investigate whether including this information in our modeling 
through a prior on $\Omega$ and $\log g$ (similar to the one used in \citealt{2015ApJ...807..183L}) can help correct systematic behavior shown by BT-Settl, where the derived posteriors frequently cluster near $T_\mathrm{eff}$ of 1800 K and $\log g$ of 5.5 dex, implying large, non-physical masses. Specifically, we follow the procedure outlined in \rfsecl{atmosphericanalysis} but place a uniform prior on $M=gR^2/G$ so that it falls between 0 $M_\mathrm{Jup}$ and 120 $M_\mathrm{Jup}$. The resulting posteriors can be found in \rftabl{derived_massprior}.

Comparisons between the parameters returned by BT-Settl with and without this mass prior are shown in \rffigl{prior-vs-noprior-hist} and \rffigl{prior-vs-noprior-scatter}. Both objects with spectral types L4.5 and later and spectral types M7.5 and earlier are hardly affected by the mass prior. In contrast, many objects with spectral types ranging from M9 to L4 jump across the parameter space under the mass prior, populating the gap in $T_\mathrm{eff}$ between 1800 K and 2500 K while the high-surface gravity objects move to lower $\log g$. However, most of these objects have mass posteriors pinned against the upper limit of 120 $M_\mathrm{Jup}$ and have $\epsilon_\mathrm{med}$ values elevated by as much as $5\%$, indicating that the model spectra fit the data much worse when using the mass prior. Consequently, we find that forward-modeling spectra with an informative prior on mass fails to yield more accurate physical parameters or models that closely fit the observed spectra for objects near the M-L transition boundary.

\begin{figure}
    \centering
    \includegraphics[width=\linewidth]{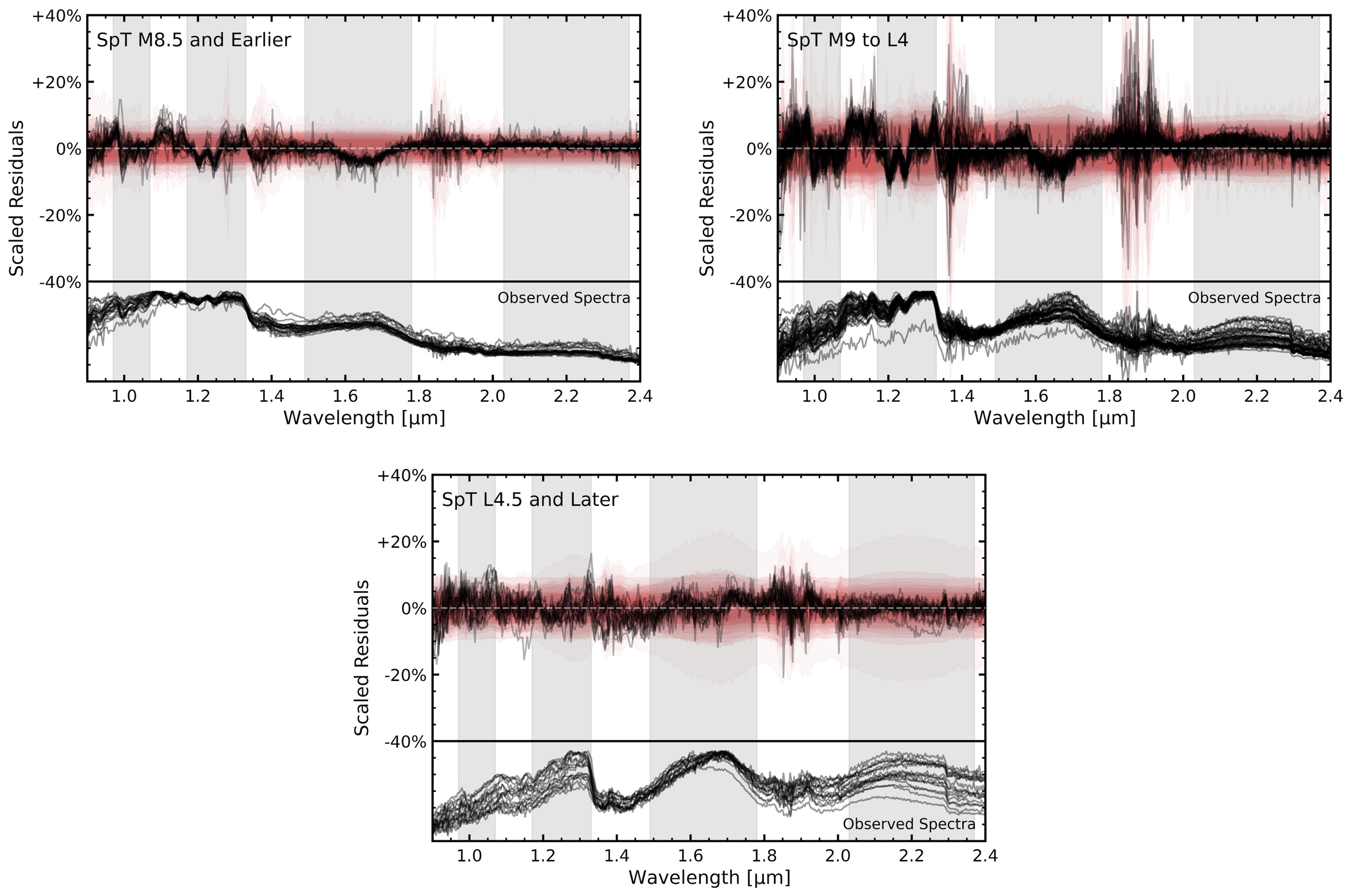}
    \caption{Stacked spectral-fitting residuals for models including interstellar extinction (using the median posterior values, normalized by the peak observed flux, and shown as black lines) for objects with spectral types M8.5 and earlier (top left panel), objects with spectral types ranging from M9 to L4 (top right panel), and those with spectral types L4.5 and later (bottom panel). The $1\sigma$ and $2\sigma$ dispersions (corresponding to the flux uncertainties and $\epsilon_\mathrm{med}$ terms added in quadrature) have been plotted for each object as red shadows. The normalized spectra for objects in each spectral type bin have been plotted at the bottom of all three panels as a reference. The wavelength intervals corresponding to the $Y$, $J$, $H$, and $K$ bands are marked by grey regions (the boundaries for each band are given in Equations \ref{eq:qY}, \ref{eq:qJ}, \ref{eq:qH}, and \ref{eq:qK}.}
    \label{fig:stackedresiduals_ext}
\end{figure}

\begin{figure}
    \centering
    \includegraphics[width=\linewidth]{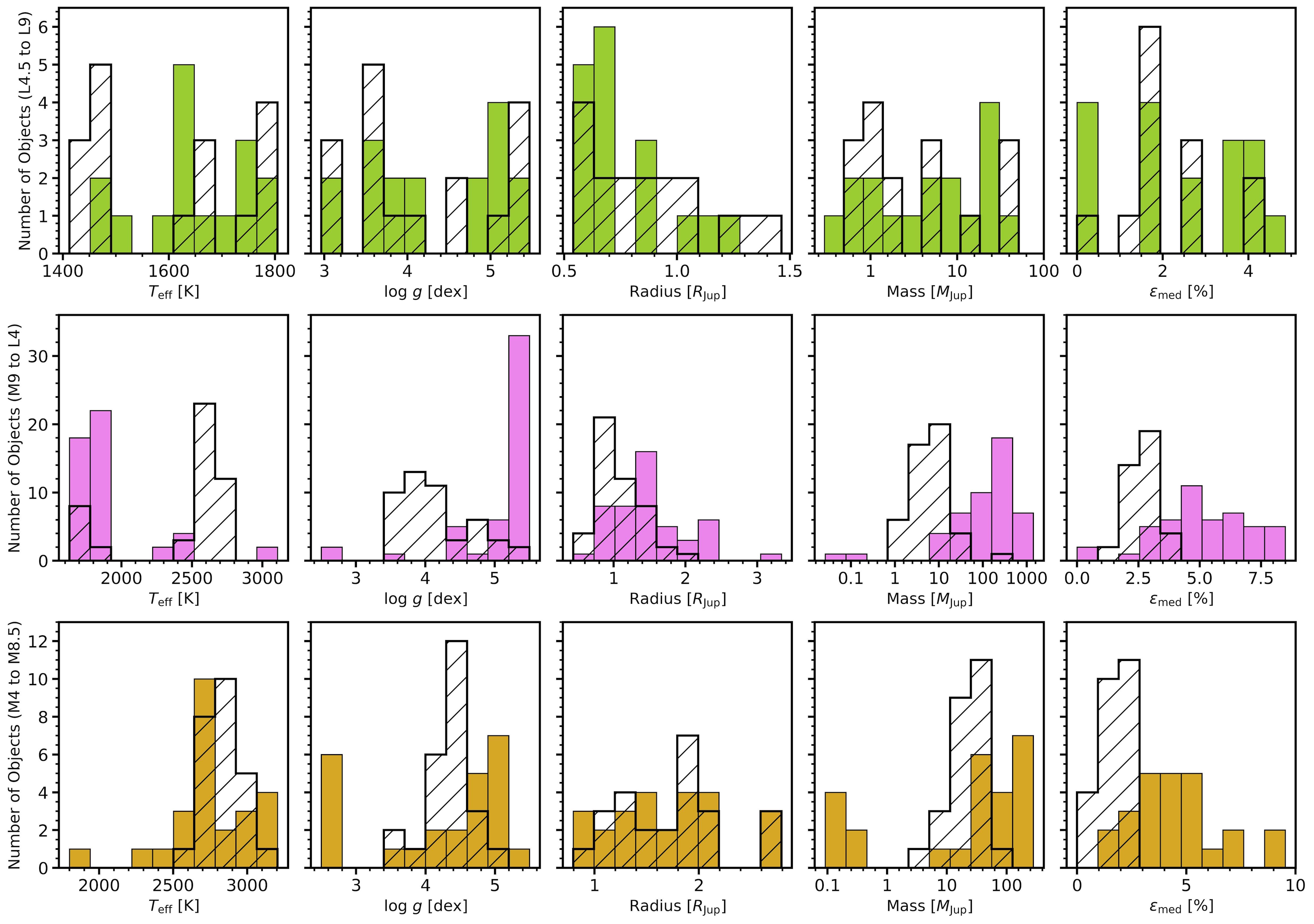}
    \caption{Histograms of parameters returned by BT-Settl without interstellar extinction (filled, colored bars) and with interstellar extinction as a free parameter (hatched, colorless bars). The top row corresponds to objects with spectral types L4.5 and later, the middle row spectral types M9 to L4, and the bottom row M8.5 and earlier.}
    \label{fig:ext-vs-noprior-hist}
\end{figure}

\begin{figure}
    \centering
    \includegraphics[width=\linewidth]{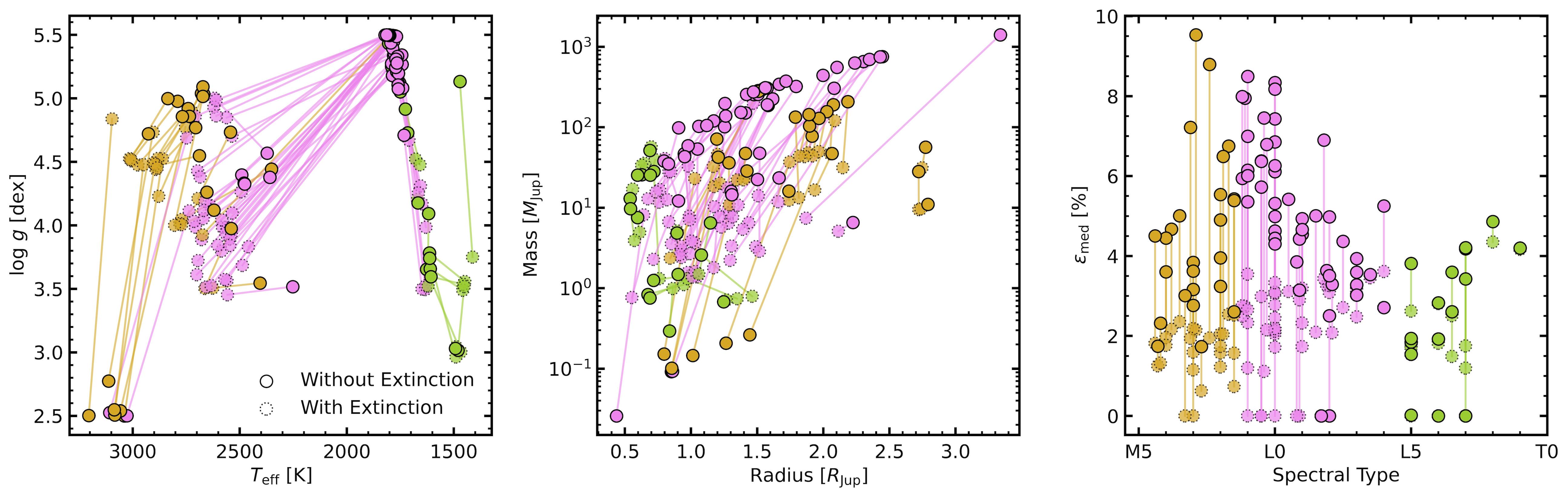}
    \caption{Points with solid edges represent the median parameters returned by BT-Settl without interstellar extinction while transparent points with dashed represent the median parameters returned with interstellar extinction as a free parameter. Lines connect points corresponding to the same object. Orange points represent objects with spectral types M8.5 and earlier, pink spectral types M9 to L4, and green L4.5 and later.}
    \label{fig:ext-vs-noprior-scatter}
\end{figure}

\begin{figure}
    \centering
    \includegraphics[width=\linewidth]{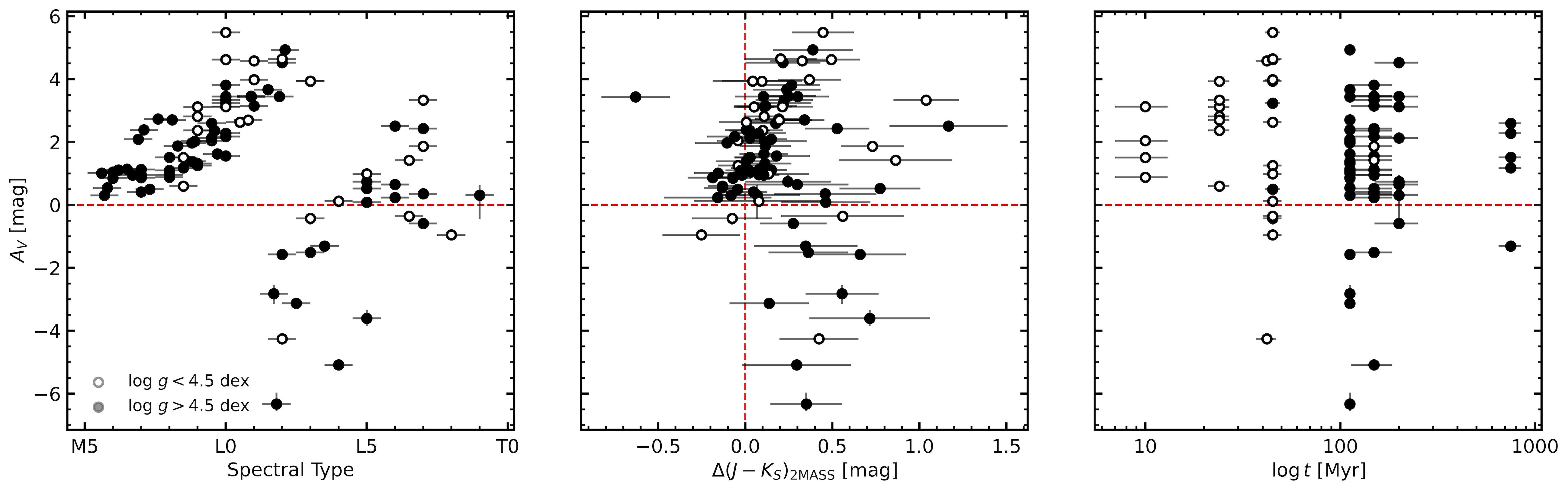}
    \caption{The extinction parameter, $A_V$, as a function of spectral type (left panel), the color anomaly ($\Delta(J-K_S)_\mathrm{2MASS}$, middle panel), and age (right panel). Filled data points correspond to objects with evolutionary surface gravities above 4.5 dex while empty data points correspond to objects with evolutionary surface gravities below 4.5 dex.}
    \label{fig:av-vs-params}
\end{figure}

\section{Improving Spectral Fits by Incorporating an Interstellar Extinction Term}
\label{sec:extinction}

\cite{2020A&A...633A.124P} fit BT-Settl model atmospheres to medium-resolution spectra of 9 objects with spectral types ranging from M8 to M9, finding residual structure similar to that described for objects near the M-L transition boundary in \rfsecl{residuals} when fitting the $J$, $H$, and $K$ bands simultaneously. However, when including interstellar extinction as a free parameter, they saw a significant reduction in their residuals across all wavelengths. We test whether accounting for reddening due to dust can improve the models and derived parameters returned by BT-Settl, following the procedure outlined in \rfsecl{atmosphericanalysis} but including the \cite{2007ApJ...663..320F} extinction law, where the extinction parameter ($A_V$) determines the amount of reddening at a given wavelength and is allowed to vary freely. We use this reddening law to model the effect of dust grains in object's atmosphere on its emergent spectra rather than actual interstellar extinction, which is negligible for objects in our sample given their close distances \citep{2019A&A...625A.135L}. The resulting parameters are given in \rftabl{derived_ext}.

The fitted spectra and residuals for each of the objects located in the Pleiades are shown in \rffigl{allspectra_pleiades_ext} while the fitted spectra and residuals for the remaining YMG and Hyades objects can be found in \rffigl{allspectra_nopleiades_ext} and \rffigl{allresiduals_nopleiades_ext}, respectively. While we see marginal improvements in the model spectra for late L dwarfs, the residuals between the models and data are reduced significantly for the rest of the objects, particularly in the $J$ and $H$ bands. This is especially apparent when comparing the stacked residuals for models with extinction (see \rffigl{stackedresiduals_ext}) to those for models without extinction (\rffigl{stackedresiduals}). Objects with spectral types M8.5 and earlier have systematic residuals near $15\%$ of the peak observed flux without extinction; including extinction reduces these residuals to less than $10\%$ of the peak observed flux. Objects with spectral types ranging from M9 to L4 see their residuals drop from $20\%$ of the peak observed flux to just above $10\%$. Further, much of the residual structure described in \rfsecl{residuals} becomes less apparent or disappears altogether. \rffigl{ext-vs-noprior-hist} and \rffigl{ext-vs-noprior-scatter} also show that the $\epsilon_\mathrm{med}$ values for these two groups of objects drop by as much as $5\%$, confirming that including extinction can vastly improve the fit between our models and data. 

Including reddening in our model may lead to better fits if BT-Settl accounts for too little or too much extinction from dust in an object's atmosphere. Positive values of $A_V$ would indicate that more extinction from dust is needed to explain the object's spectrum, whereas negative values would suggest that less extinction is needed. In \rffigl{av-vs-params}, we see that $A_V$ is greatest for objects near the M-L transition boundary, agreeing with our interpretation that BT-Settl does not fully account for extinction from dust in these objects. This indicates that BT-Settl either underestimates the concentration of dust in the atmospheres of ultracool dwarfs near the M-L transition boundary or that model atmospheres need to consider different particle size distributions, including sub-micron dust grains \citep{2016ApJ...830...96H}; it is possible that dust condensates form at higher temperatures than expected (e.g., \citealt{2020A&A...634A..23W}) and observed (e.g., \citealt{2022MNRAS.513.5701S}). While $A_V$ is negative for several objects with spectral types L2 or later, the fits between the models and data marginally improve and the derived parameters for these objects do not change drastically. Notably, $A_V$ is greatest for abnormally red objects near the M-L transition boundary, agreeing with our previous observation that BT-Settl struggles to replicate the spectra of these objects. Additionally, we do not see any correlation between gravity, age, and the extinction coefficients.

\begin{figure}
    \centering
    \includegraphics[width=\linewidth]{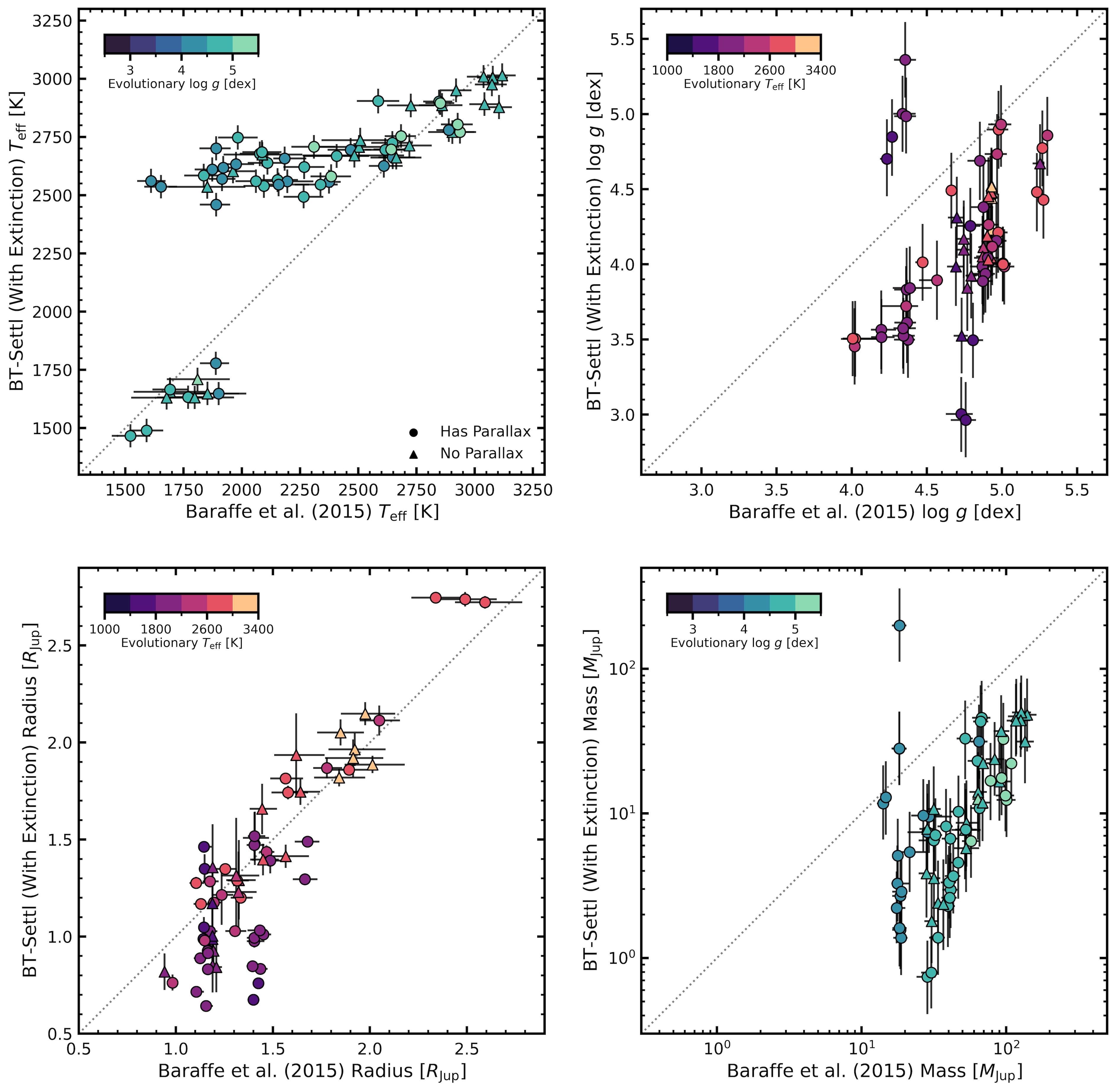}
    \caption{Physical parameters derived using BT-Settl model atmospheres when including interstellar extinction versus those derived using the \cite{2015A&A...577A..42B} evolutionary models. The top left panel compares $T_\mathrm{eff}$, the top right compares $\log g$, the bottom left $R$, and the bottom right $M$. The points in the top left and bottom right panels are colored according to the surface gravity determined using the evolutionary model while the points in the top right and bottom left are colored according to effective temperature determined using the evolutionary model. Circular data points indicate an object that has a parallax measurement while triangular data points indicate otherwise.}
    \label{fig:btsettl-vs-baraffe-ext}
\end{figure}

\begin{figure}
    \centering
    \includegraphics[width=\linewidth]{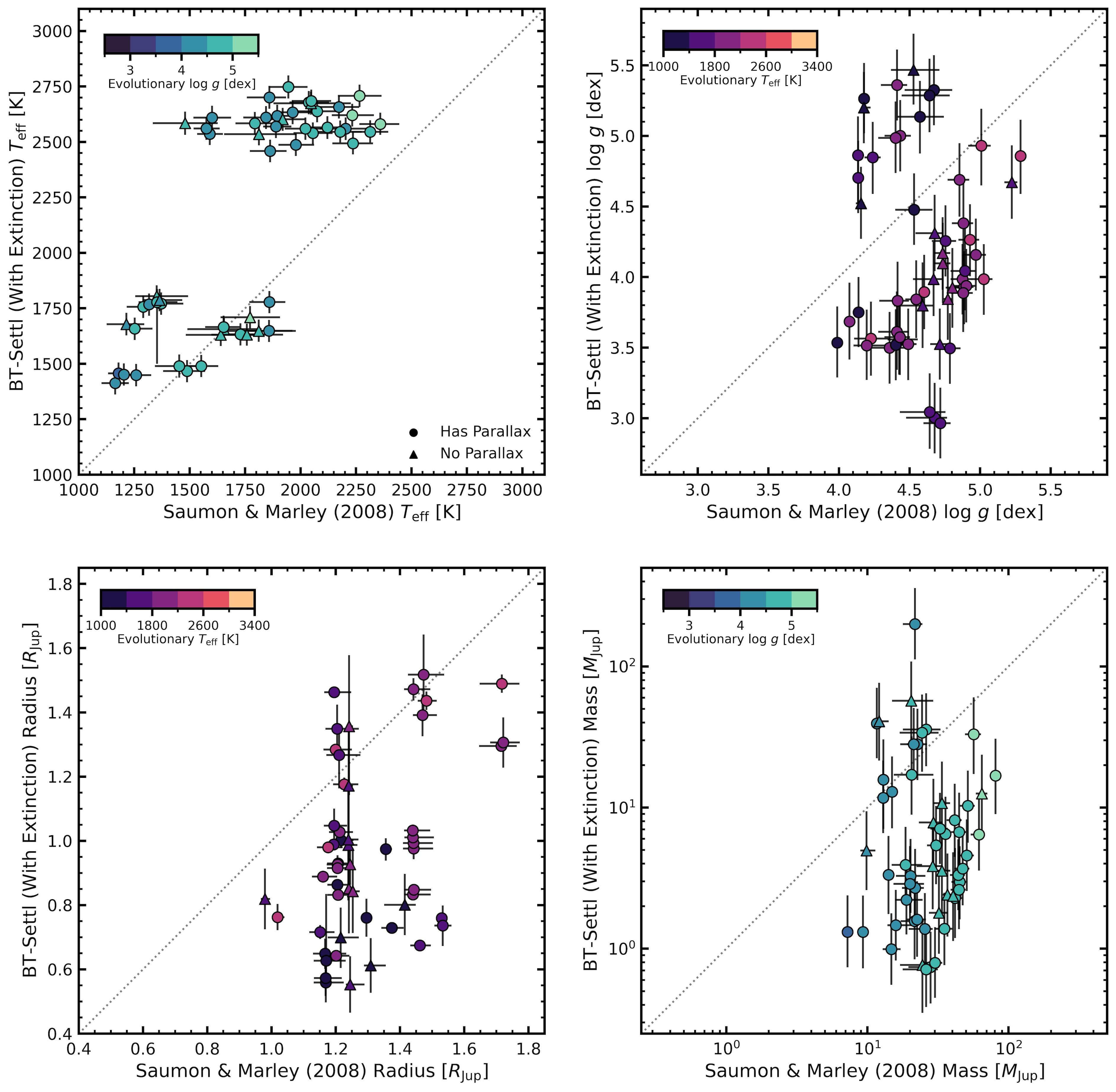}
    \caption{Physical parameters derived using BT-Settl model atmospheres when including interstellar extinction versus those derived using the \cite{2008ApJ...689.1327S} evolutionary models. The top left panel compares $T_\mathrm{eff}$, the top right compares $\log g$, the bottom left $R$, and the bottom right $M$. The points in the top left and bottom right panels are colored according to the surface gravity determined using the evolutionary model while the points in the top right and bottom left are colored according to effective temperature determined using the evolutionary model. Circular data points indicate an object that has a parallax measurement while triangular data points indicate otherwise.}
    \label{fig:btsettl-vs-sm08-ext}
\end{figure}

As seen in \rffigl{ext-vs-noprior-hist} and \rffigl{ext-vs-noprior-scatter}, the physical parameters derived from BT-Settl change drastically when including extinction. Some of these changes appear to be for the better, with the derived $\log g$ for objects near the M-L boundary shifting away from the extremes of the model grid. Correspondingly, nearly all of the masses derived when including extinction fall below the expected limit of 120 $M_\mathrm{Jup}$. However, the effective temperatures continue to be plagued with systematics, this time clustering around 1700 K and 2600 K. Comparing these parameters to those derived using evolutionary models (see \rffigl{btsettl-vs-baraffe-ext} and \rffigl{btsettl-vs-sm08-ext}) further illustrates this systematic behavior. It is possible that the parameters derived when including extinction are unreliable because we use interstellar extinction models, which account for dust with physical properties different than the dust found in ultracool atmospheres (e.g., particle size distribution, spatial distribution, and chemical composition). Perhaps fitting BT-Settl spectra with extinction models specific to the dust found in ultracool dwarf atmospheres (e.g., \citealt{2016ApJ...830...96H} for L0-L6 dwarfs and \citealt{2014MNRAS.439..372M} for an L7 dwarf) could return more accurate parameters; however, testing this falls beyond the scope of this work.

\vspace{-4mm}

\section{Discussion and Conclusions}
\label{sec:conclusions}

We fit BT-Settl model atmospheres to low-resolution ($R\approx50-250$) near-infrared (0.9 $\mathrm{\mu m}-2.5$ $\mathrm{\mu m}$) spectra of 90 late-M and L dwarfs, including 61 objects located in young moving groups and 23 objects located in the Pleiades. Comparing the physical parameters derived using atmospheric models to those found using the more robust cloudy \cite{2008ApJ...689.1327S} hybrid and \cite{2015A&A...577A..42B} evolutionary models, we find that BT-Settl returns $T_\mathrm{eff}$ clustered near 1800 K and $\log g$ pressed against the upper boundary of the model grid (5.5 dex) for relatively red objects near the M-L spectral type boundary. These high surface gravities imply impossibly large masses (150-1400~$M_\mathrm{Jup}$). In contrast, BT-Settl returns elevated $T_\mathrm{eff}$ above 3000 K and very low $\log g$ ($\lesssim$3.0 dex) for objects with typical colors near the M-L boundary. These low surface gravities imply non-physical masses of $0.02-3~M_\mathrm{Jup}$. Attempts to correct this systematic behavior through an informative prior on the mass were unsuccessful, with derived masses pinned against the limits of the prior.

The wavelengths of our spectra ($.9-2.4~\mathrm{\mu m}$) are predominantly sensitive to the molecular absorption  bands of $\mathrm{H_2O}$, which primarily carves out the $J$, $H$, and $K$ bands. BT-Settl utilizes the $\mathrm{H_2O}$ line list from \cite{Barber_2006}, which is also used in the latest ultracool atmosphere models \citep{Phillips_2020, Marley_2021} and is generally considered robust when modeling low-resolution near-infrared spectra of ultracool dwarfs. We see large residuals appear for objects near the M-L spectral type boundary, with BT-Settl consistently overpredicting the peak flux in the $J$ and $H$ bands while the overall shape of the absorption  bands are well-replicated, indicating that the $\mathrm{H_2O}$ opacity is not a main cause of uncertainty in the model atmospheres. Instead, since the peak fluxes originate from the deepest, warmest, high-pressure layers of the atmosphere, the overpredicted flux in the $J$ and $H$ bands is likely due to deficiencies in the cloud model used. BT-Settl calculates the dust grain number density and size distribution by comparing the timescales of condensation and sedimentation to mixing timescales, which are based on the 2D hydrodynamical simulations of \cite{Freytag_2010} that explore how mixing mechanisms such as convective overshoot and gravity waves keep dust aloft in the atmosphere. The overprediction of flux in the $J$ and $H$ bands for objects near the M-L transition boundary suggests that these timescales may be incorrect or not valid across the entire parameter space of late-M and L dwarfs.

Including interstellar extinction models in our spectral fitting vastly improves the fit between our models and observed spectra for late-M and early-L ultracool dwarfs, with much of the residual structure in the $J$ and $H$ bands disappearing. Extinction coefficients are greatest for objects near the M-L boundary, further implying that BT-Settl accounts for too little dust opacity in these ultracool dwarf atmospheres. The parameters derived when including interstellar extinction continue to show systematic errors, with $T_\mathrm{eff}$ clustering near 2600 K for the objects at the M-L boundary. This systematic behavior may originate from the use of an interstellar extinction model, as the properties of dust present in ultracool dwarf atmospheres are likely different from interstellar dust and would vary across spectral types. Extinction laws that account for sub-micron dust grains that are neglected by BT-Settl, such as the one presented in \cite{2016ApJ...830...96H} for early L-dwarfs, may be more appropriate.

In addition to deficiencies in the cloud model, the overprediction of flux in the $H$-band may be in part due to missing FeH opacity. Line lists for the FeH molecule commonly contain the vibrational band of the electronic transition responsible for the $0.99~\mathrm{\mu m}$ Wing-Ford band present in the near-infrared spectra of late-M and L dwarfs. However, many line lists omit the FeH transition at $1.58~\mathrm{\mu m}$. \cite{Hargreaves_2010} present a line list including this transition and is included in the latest Sonora atmospheric models \citep{Marley_2021} but contains many uncertainties, with the line strengths needing to be scaled to provide reasonable comparisons to observations of late-M and early-L dwarfs. BT-Settl does not appear to include this $1.58~\mathrm{\mu m}$ FeH band, potentially contributing to overpredicted fluxes in the $H$-band.

This analysis focused on objects located in young moving groups and the Pleiades, allowing us to benchmark the performance of model atmospheres against more robust evolutionary models. Similar to \cite{2021ApJ...921...95Z}, additional analyses should apply our forward-modeling framework to field ultracool dwarfs both to characterize these objects and better understand systematic errors in model atmospheres. Additionally, future model atmospheres that include clouds will need to better account for dust in the atmospheres of objects near the M-L spectral type transition. In the meantime, analyses that apply extinction models specific to ultracool dwarfs may mitigate systematic errors in models fitted using BT-Settl.


\begin{acknowledgments}
The authors thank the referee for thorough comments that improved the quality of the paper. The authors also thank Aaron Do, Michael Bottom, Robert Jedicke, Benjamin Shappee, Xudong Sun, and Simon Petrus for their helpful discussions and feedback. 
S.A.H. acknowledges support from the Research Experience for Undergraduates program at the University of Hawaii Institute for Astronomy funded through NSF grant 2050710 and thanks University of Hawaii for their hospitality. 
Z. Z. acknowledges support from the NASA Hubble Fellowship grant HST-HF2-51522.001-A.
This material is based upon work supported by the National Science Foundation Graduate Research Fellowship under Grant No. 2236419.
This work benefited from the 2022 Exoplanet Summer Program in the Other Worlds Laboratory (OWL) at the University of California, Santa Cruz, a program funded by the Heising-Simons Foundation. 
We thank Michael Kotson for assistance with the SpeX spectroscopy.
Finally, the authors wish to recognize and acknowledge the very significant cultural role and reverence that the summit of Maunakea has always had within the indigenous Hawaiian community. We are most fortunate to have the opportunity to conduct observations from this mountain.
\end{acknowledgments}

\software{{\tt AstroPy} \citep{astropy:2018}, {\tt Matplotlib} \citep{matplotlib}, {\tt NumPy} \citep{numpy}, {\tt SciPy} \citep{2020SciPy-NMeth}, {\tt Starfish} \citep{2015ApJ...812..128C, ian_czekala_2018_2221006}, {\tt UltraNest} \citep{2016S&C....26..383B, 2019PASP..131j8005B, 2021JOSS....6.3001B}}

\bibliography{refs}
\bibliographystyle{aasjournal}

\clearpage

\begin{longrotatetable}
\begin{deluxetable*}{@{\extracolsep{0pt}}lccccccccccccccc}
\tablecaption{Sample of M and L Dwarfs \label{tab:sample}}
\tabletypesize{\scriptsize}
\tablehead
{
\colhead{}& \colhead{}& \colhead{}& \colhead{}& \colhead{}& \colhead{}& \colhead{}& \colhead{}& \colhead{}& \colhead{}& \colhead{}& \multicolumn{5}{c}{References} \\
\cline{11-16}
\colhead{Object}& \colhead{SpT}& \colhead{Grav.}& \colhead{R.A.\tablenotemark{a}}& \colhead{Dec.\tablenotemark{a}}& \colhead{$\mu_\alpha$}& \colhead{$\mu_\delta$}& \colhead{Dist.}& \colhead{RV}& \colhead{Prob.}& \colhead{Disc.}& \colhead{SpT}& \colhead{PM}& \colhead{Dist.}& \colhead{RV} \\[-1.5ex]
\colhead{}& \colhead{}& \colhead{}& \colhead{($\degree$)}& \colhead{($\degree$)}& \colhead{($\mathrm{mas\,yr^{-1}}$)}& \colhead{($\mathrm{mas\,yr^{-1}}$)}& \colhead{(pc)}& \colhead{($\mathrm{km\,s^{-1}}$)}& \colhead{}& \colhead{}& \colhead{}& \colhead{}& \colhead{}& \colhead{}
}
\startdata
\multicolumn{16}{c}{ABDMG\tablenotemark{b}} \\
\hline
1RXS J235133.3+312720 B & L0.0 & \nodata & 357.8913 & 31.4564 & $106.58\pm0.06$ & $-46.83\pm0.04$ & $43.07\pm0.09$ & $-13.5\pm0.6$ & $0.99^{\dagger\dagger\dagger}$ & 1 & 1 & 2 & 2 & 3 \\ 
2MASS J00192626+4614078 & M8.0 & INT-G & 4.8594 & 46.2355 & $122.44\pm0.19$ & $-46.83\pm0.11$ & $38.33\pm0.25$ & $-20\pm3$ & $0.99^{\dagger\dagger\dagger}$ & 4 & 4 & 2 & 2 & 5 \\ 
2MASS J03552337+1133437 & L5.0 & VL-G & 58.8474 & 11.5622 & $219.8\pm1.6$ & $-46.8\pm0.8$ & $9.12\pm0.06$ & $11.92\pm0.22$ & $1.0^{\dagger\dagger\dagger}$ & 6 & 7 & 2 & 2 & 8 \\ 
2MASS J04203904+2355502 & L1.0 & INT-G & 65.1626 & 23.9306 & $45.1\pm1.7$ & $-46.8\pm1.2$ & $38.8\pm1.2$ & \nodata & $0.99^{\dagger\dagger}$ & 9 & 9 & 2 & 2 & \nodata \\ 
2MASS J05445741+3705039 & L1.9 & INT-G & 86.2392 & 37.0845 & $6.6\pm0.9$ & $-46.8\pm0.8$ & $29.3\pm0.5$ & \nodata & $1.0^{\dagger\dagger}$ & 10 & 10 & 2 & 2 & \nodata \\ 
2MASS J09312823+0528223 & M7.0 & \nodata & 142.8676 & 5.4729 & $-126.6\pm0.5$ & $-46.8\pm0.4$ & $28.32\pm0.14$ & 25.0 & $0.98^{\dagger\dagger\dagger}$ & 11 & 11 & 2 & 2 & 12 \\ 
2MASS J09474477+0224327 & M8.0 & \nodata & 146.9366 & 2.4091 & $-107.3\pm0.4$ & $-46.8\pm0.4$ & $33.5\pm0.28$ & \nodata & $0.98^{\dagger\dagger}$ & 11 & 11 & 2 & 2 & \nodata \\
\enddata
\end{deluxetable*}
\end{longrotatetable}

\begin{deluxetable*}{@{\extracolsep{1pt}}lccccc}
\tablecaption{Properties of Young Moving Groups\label{tab:ymg}}
\tabletypesize{\footnotesize}
\tablehead
{
\colhead{Name}& \colhead{Abbreviation}& \colhead{Distance Distribution}& \colhead{Radial Velocity Distribution}& \colhead{Age}& \colhead{Age Reference} \\[-1.5ex]
\colhead{}& \colhead{}& \colhead{(pc)}& \colhead{($\mathrm{km\,s^{-1}}$)}& \colhead{(Myr)}& \colhead{}
}
\startdata
AB Doradus Moving Group & ABDMG & $30^{+20}_{-10}$ & $10^{+10}_{-20}$ & $149^{+51}_{-19}$ & 1 \\
Argus & ARG & $15-48$ & $[-10,9]$ & $30-40$ & 2 \\
$\beta$ Pictoris Moving Group & BPMG & $30^{+20}_{-10}$ & $10\pm10$ & $24\pm3$ & 1 \\
Carina-Near & CARN & $30\pm20$ & $15^{+7}_{-10}$ & $200\pm50$ & 3 \\
Columba & COL & $50\pm20$ & $21^{+3}_{-8}$ & $42^{+6}_{-4}$ & 1 \\
Hyades Cluster & HYA & $42\pm7$ & $39^{+3}_{-4}$ & $750\pm100$ & 4 \\
Pleiades Cluster & PLE & $134\pm9$ & $6\pm2$ & $112\pm5$ & 5 \\ 
Tucana-Horologium Association & THA & $46^{+8}_{-6}$ & $9^{+5}_{-6}$ & $45\pm4$ & 1 \\
TW Hya & TWA & $60\pm10$ & $10\pm3$ & $10\pm3$ & 1
\enddata
\tablecomments{This table has been adapted from \cite{2014ApJ...783..121G} and \cite{2018ApJ...856...23G}. All distance and radial velocity distributions come from these works.}
\tablerefs{(1) \cite{2015MNRAS.454..593B}, (2) \cite{2011ApJ...732...61Z}, (3) \cite{2006ApJ...649L.115Z}, (4) \cite{2015ApJ...807...24B}, (5) \cite{2015ApJ...813..108D}}
\end{deluxetable*}
\begin{longrotatetable}
\begin{deluxetable*}{@{\extracolsep{1pt}}lcccccccc}
\tablecaption{IRTF/SpeX Observing Log and SpT Comparisons\label{tab:spex_observing}}
\tabletypesize{\footnotesize}
\tablehead
{
\colhead{Object}& \colhead{SIMBAD Name}& \colhead{Observation Date}& \colhead{Slit}& \colhead{Average Airmass}& \colhead{$T_\mathrm{int}$}& \colhead{SpT}& \colhead{Optical SpT}& \colhead{NIR SpT}\\[-1.5ex]
\colhead{}& \colhead{}& \colhead{(UTC)}& \colhead{(\arcsec)}& \colhead{}& \colhead{(s)}&  \colhead{(This Work)}& \colhead{(Literature)}& \colhead{(Literature)}
}
\startdata
\multicolumn{9}{c}{ABDMG} \\
\hline
2MASS J00192626+4614078      &   PM J00194+4614                                            & 2015-07-15  &  0.5 & 1.12 &     88  & M8.0 & \nodata & \nodata \\  
2MASS J04203904+2355502      &   2MASS J04203904+2355502                                   & 2015-12-24  &  0.5 & 1.04 &    386 & L1.0 & \nodata & \nodata  \\  
2MASS J05445741+3705039      &   2MASS J05445741+3705039     & 2015-10-14  &  0.5 & 1.09 &    133  & L1.9 & \nodata & \nodata \\  
PSO J058.0039+41.7950        &   \nodata                                       & 2016-03-22  &  0.5 & 1.31 &    777 & M9.6 & \nodata & \nodata  \\  
PSO J318.4243+35.1277        &   PSO J318.4243+35.1277                                   & 2013-09-22  &  0.5 & 1.10 &     60 & L0.9 & \nodata & \nodata  \\  
\hline 
\multicolumn{9}{c}{BPMG} \\
\hline
PSO J316.1306$-$09.6562      &   2MASS J21043128-0939217               & 2015-11-28  &  0.5 & 1.29 &    837  & L0.8 & \nodata & \nodata \\
\enddata

\tablerefs{(1) \cite{1998ApJ...499L.199S}, (2) \cite{2010AA...519A..93B}, (3) \cite{2000ApJ...543..299M}, (4) \cite{1999ApJ...519..802K}, (5) \cite{1997ApJ...491L..81Z}, (6) \cite{1996ApJ...469..706M}, (7) \cite{2014AA...572A..67Z}, (8) \cite{1995MNRAS.272..630S}, (9) \cite{1998MNRAS.298L..34F}, (10) \cite{1998ApJ...499L..61M}}
\end{deluxetable*}
\end{longrotatetable}

\begin{longrotatetable}
\begin{deluxetable*}{@{\extracolsep{2pt}}lccccccccccccc}
\tablecaption{Photometry and Information about Spectra for M and L Dwarf Sample \label{tab:photometry}}
\tablewidth{700pt}
\tabletypesize{\scriptsize}
\tablehead
{
\colhead{}& \colhead{}& \multicolumn{4}{c}{MKO/UKIDSS}& \multicolumn{4}{c}{2MASS}& \multicolumn{2}{c}{Spectrum}\\
\cline{3-6} \cline{7-10} \cline{11-13}
\colhead{Object}& \colhead{SpT}& \colhead{$J$}& \colhead{$H$}& \colhead{$K$}& \colhead{Refs}& \colhead{$J$}& \colhead{$H$}& \colhead{$Ks$}& \colhead{Refs}& \colhead{$\overline{\mathrm{SNR}}$}& \colhead{Refs}
}
\startdata
\multicolumn{12}{c}{ABDMG} \\
\hline
1RXS J235133.3+312720 B & L0.0 & $15.3\pm0.1$ & $14.64\pm0.05$ & $13.93\pm0.05$ & 1,2 & $15.4\pm0.1$ & $14.57\pm0.05$ & $13.94\pm0.05$ & 2 & 85 & 2 \\ 
2MASS J00192626+4614078 & M8.0 & $12.548\pm0.021$ & $11.98\pm0.014$ & $11.47\pm0.014$ & 1 & $12.603\pm0.021$ & $11.94\pm0.014$ & $11.502\pm0.014$ & 3 & 130 & 4 \\ 
2MASS J03552337+1133437 & L5.0 & $13.94\pm0.01$ & $12.601\pm0.021$ & $11.502\pm0.001$ & 1,5,6 & $14.05\pm0.024$ & $12.53\pm0.021$ & $11.526\pm0.021$ & 3 & 258 & 7 \\ 
2MASS J04203904+2355502 & L1.0 & $14.85\pm0.03$ & $14.13\pm0.03$ & $13.476\pm0.003$ & 1,8 & $14.94\pm0.03$ & $14.08\pm0.03$ & $13.499\pm0.029$ & 3 & 42 & 4 \\ 
2MASS J05445741+3705039 & L1.9 & $13.911\pm0.002$ & $13.13\pm0.001$ & $12.386\pm0.001$ & 9 & $14.03\pm0.03$ & $13.034\pm0.033$ & $12.408\pm0.023$ & 3 & 85 & 4 \\ 
2MASS J09312823+0528223 & M7.0 & $12.754\pm0.001$ & $12.207\pm0.001$ & $11.765\pm0.001$ & 8 & $12.864\pm0.027$ & $12.226\pm0.023$ & $11.796\pm0.021$ & 3 & 193 & 7 \\ 
2MASS J09474477+0224327 & M8.0 & $13.163\pm0.002$ & $12.57\pm0.001$ & $12.088\pm0.002$ & 8 & $13.174\pm0.026$ & $12.485\pm0.024$ & $12.079\pm0.027$ & 3 & 157 & 7 \\
\enddata

\tablerefs{(1) \cite{2021AJ....161...42B}, (2) \cite{2015ApJS..216....7B}, (3) \cite{2003tmc..book.....C}, (4) This Work, (5) \cite{2018MNRAS.473.5113D}, (6) \cite{2007MNRAS.379.1599L}, (7) \cite{2014ApJ...794..143B}, (8) \cite{2012yCat.2314....0L}, (9) \cite{2012yCat.2316....0U}, (10) \cite{2017ASInC..14....7B}, (11) \cite{2016ApJS..225...10F}, (12) \cite{2004AJ....127.3553K}, (13) \cite{2002AJ....124.1170D}, (14) \cite{2013Msngr.154...35M}, (15) \cite{2012ApJ...752...56F}, (16) \cite{2002ApJ...564..452L}, (17) \cite{2015ApJ...814..118B}, (18) \cite{2015ApJ...814..118B}, (19) \cite{2016ApJ...821..120A}, (20) \cite{2014AJ....147...34S}, (21) \cite{2012AJ....144...94G}, (22) \cite{2008AJ....136.1290R}, (23) \cite{2011ApJ...732...56G}, (24) \cite{2003IAUS..211..197W}, (25) \cite{1997AA...327L..25D}, (26) \cite{2006AJ....131.2722C}, (27) \cite{2020AJ....159..257B}, (28) \cite{2002AJ....123.3409H}, (29) \cite{2013ApJ...772...79A}, (30) \cite{2013ApJ...777L..20L}, (31) \cite{2016ApJS..225...10F}, (32) \cite{2003AJ....126.2421C}, (33) \cite{2013ApJ...777...84B}, (34) \cite{2013ApJ...777...84B}, (35) Synthesized from Spectrum, (36) \cite{2014ApJ...787..126L}, (37) \cite{2013yCat.2319....0L}, (38) \cite{2006ApJ...639.1120K}, (39) \cite{2016yCat.2343....0E}, (40) \cite{2016ApJ...833...96L}, (41) \cite{2014ApJ...785L..14G}, (42) \cite{2002ApJ...575..484G}, (43) \cite{2007ApJ...669L..97L}, (44) \cite{2015ApJS..219...33G}}

\end{deluxetable*}
\end{longrotatetable}

\startlongtable
\begin{deluxetable*}{@{\extracolsep{10pt}}lcccccc}
\tablecaption{Parameters for M and L Dwarfs Derived Using Atmospheric Models\label{tab:derived_btsettl}}
\tabletypesize{\footnotesize}
\tablehead
{
\colhead{Object}& \colhead{SpT}& \colhead{$T_\mathrm{eff}$}& \colhead{$\log g$}& \colhead{$R$}& \colhead{$M$}& \colhead{$\epsilon_\mathrm{med}$} \\
\colhead{}& \colhead{}& \colhead{(K)}& \colhead{(dex)}& \colhead{($R_\mathrm{Jup}$)}& \colhead{($M_\mathrm{Jup}$)}& \colhead{($\%$)}
}
\startdata
\multicolumn{7}{c}{ABDMG} \\
\hline
1RXS J235133.3+312720 B & L0.0 & $3108^{+51}_{-53}$ & $2.53\pm0.25$ & $0.4368^{+0.0047}_{-0.0046}$ & $0.026^{+0.02}_{-0.011}$ & $6.11\pm0.26$ \\ 
2MASS J00192626+4614078 & M8.0 & $2672^{+51}_{-52}$ & $5.02\pm0.25$ & $1.789\pm0.016$ & $134^{+100}_{-58}$ & $3.95\pm0.12$ \\ 
2MASS J03552337+1133437 & L5.0 & $1613^{+50}_{-51}$ & $3.74^{+0.25}_{-0.26}$ & $1.0784\pm0.0076$ & $2.6^{+2.0}_{-1.2}$ & $1.9\pm0.1$ \\ 
2MASS J04203904+2355502 & L1.0 & $1786^{+51}_{-50}$ & $5.5^{+0.25}_{-0.26}$ & $1.508^{+0.048}_{-0.047}$ & $286^{+230}_{-130}$ & $4.54^{+0.18}_{-0.16}$ \\ 
2MASS J05445741+3705039 & L1.9 & $1786^{+51}_{-50}$ & $5.39\pm0.25$ & $1.795^{+0.031}_{-0.03}$ & $322^{+250}_{-140}$ & $3.64\pm0.11$ \\ 
2MASS J09312823+0528223 & M7.0 & $2672\pm50$ & $5.09^{+0.24}_{-0.25}$ & $1.195^{+0.0077}_{-0.0078}$ & $71^{+54}_{-31}$ & $2.763^{+0.1}_{-0.095}$ \\ 
2MASS J09474477+0224327 & M8.0 & $2543\pm51$ & $4.73^{+0.26}_{-0.25}$ & $1.286^{+0.015}_{-0.016}$ & $36^{+29}_{-16}$ & $4.9^{+0.18}_{-0.17}$ \\
\enddata
\tablecomments{Values represent the median of each posterior with $68\%$ confidence interval.}
\end{deluxetable*}

\begin{longrotatetable}
\begin{deluxetable*}{@{\extracolsep{5 pt}}lccccccccc}
\tablecaption{Parameters for M and L Dwarfs Derived Using Evolutionary Sequences\label{tab:derived_evo}}
\tabletypesize{\scriptsize}
\tablehead
{
\colhead{}& \colhead{}& \multicolumn{4}{c}{Saumon $\&$ Marley (2008) Hybrid Model}& \multicolumn{4}{c}{Baraffe et al. (2015) Model} \\
\cline{3-6} \cline{7-10}
\colhead{Object}& \colhead{SpT}&
\colhead{$T_\mathrm{eff}$}& \colhead{$\log g$}& \colhead{$R$}& \colhead{$M$}& \colhead{$T_\mathrm{eff}$}& \colhead{$\log g$}& \colhead{$R$}& \colhead{$M$} \\[-2ex]
\colhead{}& \colhead{}& \colhead{(K)}& \colhead{(dex)}& \colhead{($R_\mathrm{Jup}$)}& \colhead{($M_\mathrm{Jup}$)} & \colhead{(K)}& \colhead{(dex)}& \colhead{($R_\mathrm{Jup}$)}& \colhead{($M_\mathrm{Jup}$)}
}
\startdata
\multicolumn{10}{c}{ABDMG} \\
\hline
1RXS J235133.3+312720 B & L0.0 & $1945^{+86}_{-87}$ & $4.854^{+0.07}_{-0.084}$ & $1.201^{+0.041}_{-0.034}$ & $41.6^{+4.9}_{-5.2}$ & $1983\pm85$ & $4.854^{+0.065}_{-0.083}$ & $1.155^{+0.036}_{-0.029}$ & $38.4^{+4.4}_{-4.8}$ \\ 
2MASS J00192626+4614078 & M8.0 & \nodata & \nodata & \nodata & \nodata & $2927^{+65}_{-72}$ & $5.007^{+0.052}_{-0.078}$ & $1.566^{+0.091}_{-0.079}$ & $99^{+15}_{-14}$ \\ 
2MASS J03552337+1133437 & L5.0 & $1552^{+77}_{-71}$ & $4.718^{+0.072}_{-0.12}$ & $1.195^{+0.053}_{-0.031}$ & $30.1^{+3.8}_{-5.2}$ & $1591^{+72}_{-71}$ & $4.758^{+0.069}_{-0.099}$ & $1.145^{+0.04}_{-0.029}$ & $30.3^{+3.5}_{-4.5}$ \\ 
2MASS J04203904+2355502 & L1.0 & $2022^{+92}_{-93}$ & $4.875^{+0.069}_{-0.084}$ & $1.206^{+0.041}_{-0.034}$ & $44.0^{+5.2}_{-5.6}$ & $2061^{+89}_{-88}$ & $4.869^{+0.065}_{-0.082}$ & $1.164^{+0.037}_{-0.03}$ & $40.3^{+4.8}_{-5.1}$ \\ 
2MASS J05445741+3705039 & L1.9 & $2236^{+91}_{-93}$ & $4.928^{+0.065}_{-0.082}$ & $1.227^{+0.043}_{-0.035}$ & $51.4^{+6.3}_{-6.5}$ & $2265^{+85}_{-86}$ & $4.911^{+0.066}_{-0.083}$ & $1.196^{+0.04}_{-0.033}$ & $46.8^{+5.8}_{-6.1}$ \\ 
2MASS J09312823+0528223 & M7.0 & \nodata & \nodata & \nodata & \nodata & $2647^{+88}_{-91}$ & $4.977^{+0.065}_{-0.079}$ & $1.334^{+0.061}_{-0.052}$ & $67.7^{+10.0}_{-9.6}$ \\ 
2MASS J09474477+0224327 & M8.0 & \nodata & \nodata & \nodata & \nodata & $2616^{+82}_{-86}$ & $4.974^{+0.063}_{-0.082}$ & $1.316^{+0.06}_{-0.049}$ & $65.6\pm9.1$ \\
\enddata
\tablecomments{Values represent the median of each posterior with $68\%$ confidence interval.}
\end{deluxetable*}
\end{longrotatetable}
\startlongtable
\begin{deluxetable*}{@{\extracolsep{10pt}}lcccccc}
\tablecaption{Parameters for M and L Dwarfs Derived Using Atmospheric Models With a Prior on Mass\label{tab:derived_massprior}}
\tabletypesize{\footnotesize}
\tablehead
{
\colhead{Object}& \colhead{SpT}& \colhead{$T_\mathrm{eff}$}& \colhead{$\log g$}& \colhead{$R$}& \colhead{$M$}& \colhead{$\epsilon_\mathrm{med}$} \\
\colhead{}& \colhead{}& \colhead{(K)}& \colhead{(dex)}& \colhead{($R_\mathrm{Jup}$)}& \colhead{($M_\mathrm{Jup}$)}& \colhead{($\%$)}
}
\startdata
\multicolumn{7}{c}{ABDMG} \\
\hline
1RXS J235133.3+312720 B & L0.0 & $3108\pm53$ & $2.52\pm0.25$ & $0.4366^{+0.0048}_{-0.0044}$ & $0.025^{+0.02}_{-0.011}$ & $6.1^{+0.25}_{-0.24}$ \\ 
2MASS J00192626+4614078 & M8.0 & $2695^{+53}_{-54}$ & $4.92^{+0.27}_{-0.31}$ & $1.752^{+0.017}_{-0.013}$ & $104^{+88}_{-53}$ & $3.98^{+0.13}_{-0.11}$ \\ 
2MASS J03552337+1133437 & L5.0 & $1612\pm50$ & $3.74^{+0.25}_{-0.26}$ & $1.0785^{+0.0075}_{-0.0074}$ & $2.6^{+2.0}_{-1.2}$ & $1.94^{+0.1}_{-0.094}$ \\ 
2MASS J04203904+2355502 & L1.0 & $1871^{+51}_{-50}$ & $5.21\pm0.25$ & $1.353\pm0.042$ & $120^{+94}_{-53}$ & $7.47^{+0.25}_{-0.23}$ \\ 
2MASS J05445741+3705039 & L1.9 & $1990^{+49}_{-50}$ & $5.11\pm0.25$ & $1.441\pm0.026$ & $108^{+86}_{-47}$ & $7.37^{+0.23}_{-0.21}$ \\ 
2MASS J09312823+0528223 & M7.0 & $2673^{+51}_{-50}$ & $5.1\pm0.25$ & $1.1952^{+0.0078}_{-0.0079}$ & $72^{+56}_{-32}$ & $2.761^{+0.1}_{-0.094}$ \\ 
2MASS J09474477+0224327 & M8.0 & $2543\pm51$ & $4.74^{+0.26}_{-0.25}$ & $1.286\pm0.015$ & $36^{+29}_{-16}$ & $4.91^{+0.19}_{-0.17}$ \\
\enddata
\tablecomments{Values represent the median of each posterior with $68\%$ confidence interval.}
\end{deluxetable*}

\startlongtable
\begin{deluxetable*}{@{\extracolsep{2pt}}lccccccc}
\tablecaption{Parameters for M and L Dwarfs Derived Using Atmospheric Models With Interstellar Extinction\label{tab:derived_ext}}
\tabletypesize{\footnotesize}
\tablehead
{
\colhead{Object}& \colhead{SpT}& \colhead{$T_\mathrm{eff}$}& \colhead{$\log g$}& \colhead{$R$}& \colhead{$M$}& \colhead{$A_V$}& \colhead{$\epsilon_\mathrm{med}$} \\
\colhead{}& \colhead{}& \colhead{(K)}& \colhead{(dex)}& \colhead{($R_\mathrm{Jup}$)}& \colhead{($M_\mathrm{Jup}$)}& \colhead{(mag)}& \colhead{($\%$)}
}
\startdata
\multicolumn{8}{c}{ABDMG} \\
\hline
1RXS J235133.3+312720 B & L0.0 & $2748\pm52$ & $4.69\pm0.26$ & $0.6425^{+0.0048}_{-0.0052}$ & $8.1^{+6.6}_{-3.7}$ & $1.561^{+0.052}_{-0.046}$ & $2.45^{+0.24}_{-0.22}$ \\ 
2MASS J00192626+4614078 & M8.0 & $2804\pm50$ & $4.0\pm0.25$ & $1.815\pm0.013$ & $13.3^{+10.0}_{-5.8}$ & $0.949\pm0.018$ & $1.573^{+0.069}_{-0.061}$ \\ 
2MASS J03552337+1133437 & L5.0 & $1490\pm50$ & $2.96\pm0.25$ & $1.46\pm0.01$ & $0.79^{+0.63}_{-0.35}$ & $0.522^{+0.046}_{-0.045}$ & $1.71\pm0.086$ \\ 
2MASS J04203904+2355502 & L1.0 & $2560\pm50$ & $3.98\pm0.25$ & $0.926\pm0.029$ & $3.3^{+2.6}_{-1.5}$ & $3.147^{+0.035}_{-0.034}$ & $2.32^{+0.15}_{-0.13}$ \\ 
2MASS J05445741+3705039 & L1.9 & $2493\pm49$ & $4.26^{+0.25}_{-0.26}$ & $1.176\pm0.021$ & $10.3^{+8.0}_{-4.7}$ & $3.444^{+0.036}_{-0.033}$ & $3.27\pm0.11$ \\ 
2MASS J09312823+0528223 & M7.0 & $2725\pm50$ & $4.9\pm0.26$ & $1.2007^{+0.0077}_{-0.0075}$ & $46^{+37}_{-21}$ & $0.411^{+0.025}_{-0.026}$ & $2.182^{+0.082}_{-0.078}$ \\ 
2MASS J09474477+0224327 & M8.0 & $2695^{+51}_{-50}$ & $4.21^{+0.26}_{-0.27}$ & $1.289\pm0.012$ & $10.9^{+8.9}_{-5.0}$ & $1.1^{+0.028}_{-0.026}$ & $2.049^{+0.073}_{-0.07}$ \\
\enddata
\tablecomments{Values represent the median of each posterior with $68\%$ confidence interval.}
\end{deluxetable*}

\end{document}